\pdfoutput=1 
\documentclass{iopart}

\usepackage{svn-multi}
\svnidlong 
{$HeadURL: file:///Users/ojennric/SVN/Paper/LISA_Technology/main.tex $} 
{$LastChangedDate: 2009-06-15 11:26:41 +0200 (Mon, 15 Jun 2009) $} 
{$LastChangedRevision: 50 $} 
{$LastChangedBy: ojennric $} 
\svnid{$Id: main.tex 50 2009-06-15 09:26:41Z ojennric $} 

\usepackage[latin1]{inputenc}
\usepackage[T1]{fontenc}

\usepackage{graphicx}
\usepackage{tikz}
\usepackage[log,noload=abbr]{siunitx}
\usepackage{textcomp}
\usepackage{tabularx}
\usepackage{array}
\usepackage{dcolumn}

\usepackage[%
bibtex8=true%
,bibstyle=IOP%
,citestyle=numeric-comp%
,sorting=none%
,firstinits=true%
,terseinits=true%
]{biblatex}

\usepackage{LISA_tgif}

\newunit[per=slash]{\rthz}{\ensuremath{\sqrt\text{Hz}}}
\newunit{\one}{1}
\newunit{\msol}{\ensuremath{M_\odot}}

\newcommand{\eg}{e.\,g.}
\newcommand{\ie}{i.\,e.}
\newcommand{\pll}{\parallel}   
\newcommand{\uff}[1][3]{\sqrt{1+\left(\frac{\SI{#1}{mHz}}{f}\right)^4}} 

\setkeys{Gin}{width=\linewidth}

\usetikzlibrary{arrows,matrix,chains,shapes,backgrounds,positioning,calc,fit,shadows}
\tikzset{key/.style={font=\footnotesize}}

\bibliography{all}

\begin{document}
\topical{LISA technology and instrumentation}
\author{O Jennrich$^1$}
\address{ESA/ESTEC, Keplerlaan 1, 2200 AG Noordwijk, The Netherlands}
\ead{oliver.jennrich@esa.int}

\begin{abstract}
  This article reviews the present status of the technology and
  instrumentation for the joint ESA/NASA gravitational wave detector
  LISA. It briefly describes the measurement principle and the mission
  architecture including the resulting sensitivity before focussing on
  a description of the main payload items, such as the interferomtric
  measurement system, comprising the optical system with the optical
  bench and the telescope, the laser system, and the phase measurement
  system; and the disturbance reduction system with the inertial
  sensor, the charge control system, and the micropropulsion
  system. The article touches upon the requirements for the different
  subsystems that need to be fulfilled to obtain the overall
  sensitivity.
\end{abstract}
\pacs{%
04.80.Nn 
,95.55.Ym 
,07.87.+v %
,07.60.Ly 
}

\submitto{\CQG}

\maketitle

\svnidlong 
{$HeadURL: file:///Users/ojennric/SVN/Paper/LISA_Technology/Overview.tex $} 
{$LastChangedDate: 2009-06-15 16:43:52 +0200 (Mon, 15 Jun 2009) $} 
{$LastChangedRevision: 55 $} 
{$LastChangedBy: ojennric $} 
\svnid{$Id: Overview.tex 55 2009-06-15 14:43:52Z ojennric $} 

\section{Introduction}
\label{sec:lisa-miss-overv}

LISA is a space mission, jointly planned by ESA and NASA with the
purpose to detect and observe low frequency gravitational waves
\cite{johann_european_2008,stebbins_lisa_2006,danzmann_lisa_2003,bender_lisa_1996}. In
contrast to ground-based gravitational wave detectors
\cite{hough_long_2007,sigg_status_2006,willke_geo600:_2007,acernese_status_2006,shoemaker_ground-based_2003,ando_current_2002,robertson_laser_2000}
that have a typical sensitivity in the range from \SI{1}{\hertz} to \SI{1}{\kilo\hertz},
the sensitivity for LISA stretches between \SI{0.1}{\milli\hertz} and
\SI{0.1}{\hertz}, accessing a frequency window that is inaccessible to
ground-based detectors due to seismic noise and gravity gradient
noise. 

The sensitivity for low frequencies allows LISA to to assess
gravitational waves that are emitted by some of the most violent
events in the Universe, such as the coalescence of massive black holes
($m_\text{BH} \approx \SI{e5}{\msol} \ldots \SI{e7}{\msol}$). Other
sources include the capture of stellar-size compact objects by massive
black holes and the signal from binary systems in our galaxy. The
science of LISA is discussed in detail in
\cite{menou_cosmological_2008,macleod_precision_2008,deffayet_probing_2007,amaro-seoane_topical_2007,hogan_gravitational_2006,holz_using_2005,berti_testing_2005}
and is beyond the scope of this paper.

\begin{figure}
  \centering
  \includegraphics{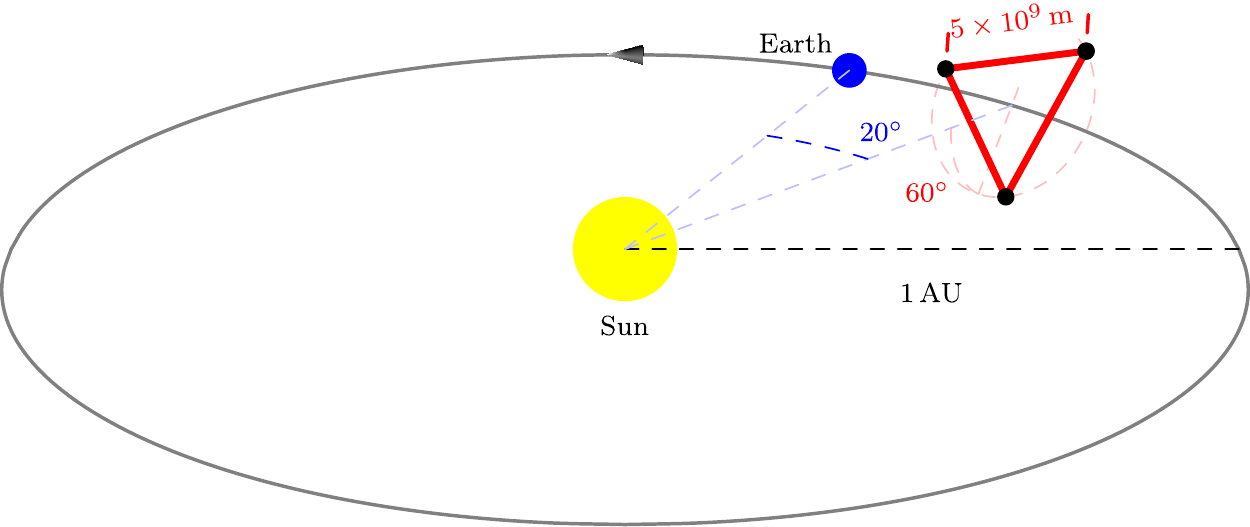}
  \caption{Schematics of the LISA constellation orbiting the Sun. The
    constellation is inclined by \ang{60} with respect to the ecliptic
    and trails the Earth by about \ang{20}, resulting in a distance of
  about \SI{50e6}{\kilo\meter} to Earth. The distance between the
  satellites (the armlength) is nominally \SI{5e6}{\kilo\meter}.}
  \label{fig:LISA_orb_scheme}
\end{figure}

\subsection{Mission concept}
\label{sec:mission-concept}

LISA comprises three spacecraft in a heliocentric orbit, forming an
equilateral triangle with \SI{5e6}{\kilo\meter} a side (the
``armlength'' of the constellation), inclined by \ang{60} with respect
to the ecliptic and trailing the Earth by about \ang{20}
(see~\fref{fig:LISA_orb_scheme}). 
Basic orbital mechanics \cite{folkner_lisa_1997} cause the
constellation to maintain its size and shape closely for the nominal
duration of the mission, allowing to operate LISA without further
station keeping manoeuvres. However, tidal forces on the constellation
cause the distances between the satellites to change
by up to \SI{1}{\percent}, resulting in a differential velocity along the
line of sight of up to \SI[per=slash]{15}{\meter\per\second} and in 
slight change of the constellation's shape
(\fref{fig:const_char}). The distance toward the Earth, approximately
\SI{5e7}{km}, has been chosen as a compromise between
long-term stability of the constellation and communications
requirements.

To reach the operational orbits, the three spacecraft are provided
with propulsion modules that will allow manoeuvering during the 14
month cruise phase and the final orbit insertion, after which the
propulsion modules will be jettisoned.

The nominal mission life is 5 years, the launch is currently foreseen
in the 2020 timeframe.

As gravitational waves cause a strain in the space-time
\cite{misner_gravitation_1973}, measuring the change in the distance
of spatially separated objects is a common concept for measuring the
effect of gravitational waves.  Each spacecraft houses two test
masses, kept as close as possible in free fall, that form the
reference points for an interferometric measurement of the
inter-spacecraft distance.

The ability to keep the test masses in near geodesic motion, the
required control law, aspects of the interferometry, and the \textmu
N-thrusters as well as the operational aspects of such a mission will
be demonstrated in the LISA Pathfinder mission, a technology precursor
to LISA that is foreseen to launch in 2011. LISA Pathfinder will put
two test masses in a near-perfect gravitational free fall, and control
and measure their motion with an accuracy that is only factors of a
few less than required for LISA. While the hardware of LISA pathfinder
has been designed with the LISA performance in mind, in order to save
on testing and verification, the requirements on the acceleration
noise for LISA Pathfinder and the lower end of the frequency band are
about a factor of 10 relaxed compared to LISA.

 A full description of LISA Pathfinder
is beyond the scope of this paper and the reader is referred to
\cite{mcnamara_lisa_2008,2009CQGra..26i4001A} and references therein.

\subsection{Sensitivity}
\label{sec:sensitivity}

\begin{figure}
  \centering
  {
    \footnotesize
    \setkeys{Gin}{width=5in}
    \input{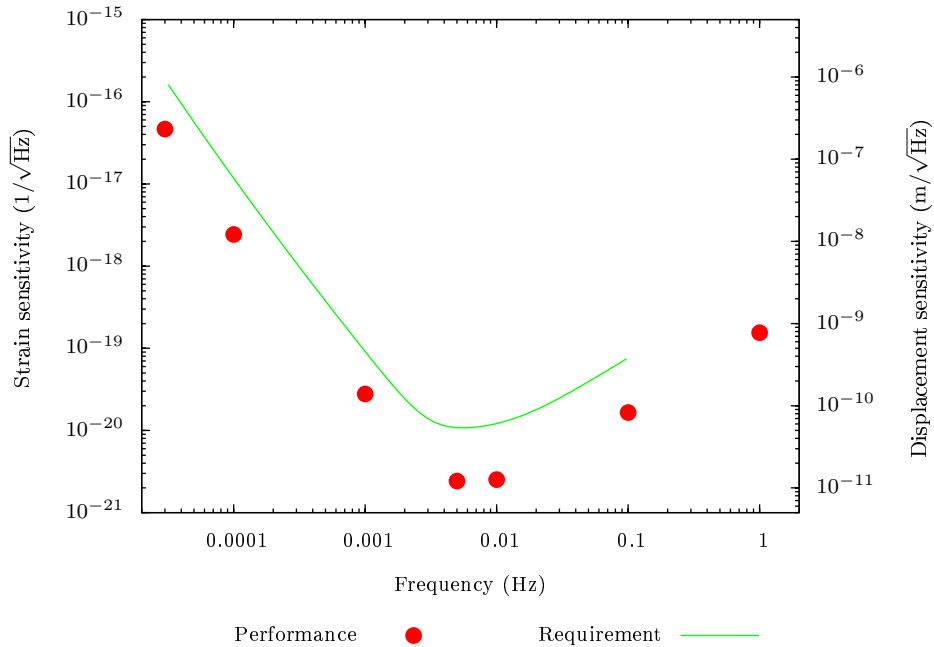}
    }
    \caption{Sensitivity requirements (green solid) and current best
      estimate of the performance (red dots). The strain sensitivity
      (left axis) is given for the full instrument, the displacement
      sensitivity (right axis) has been converted to a \emph{single
        arm} link, \ie\ represents the equivalent accuracy for the
      armlength measurement, combining all noise sources. The design
      of the instrument is such that a $\sim\SI{35}{\percent}$
      performance margin is included, resulting in the values for the
      current best estimate to be consistently lower than the
      requirement.}
  \label{fig:sensitivity}
\end{figure}

 The interferometric measurement allows
to assess the distance between the (almost)
free-falling test masses to
a level of \SI[per=slash]{10}{\pico\meter\per\rthz}, resulting in a
strain sensitivity of about \SI[per=slash]{e-20}{\per\rthz}, enough to
detect, \eg\ the coalescence of massive black holes even at redshifts
of $z=20$ with a signal-to-noise-ratio of several hundreds
\cite{lang_localizing_2008,lang_measuring_2006,holz_using_2005}. The
science requirements for LISA define the sensitivity in a
frequency window from \SI{3e-5}{\hertz} to \SI{0.1}{\hertz}
(\fref{fig:sensitivity}) as this is where most of the sources for LISA
are expected to emit.


The limitation to the sensitivity at frequencies below approximately
\SI{3}{\milli\hertz} is given by residual acceleration noise of the
order of \SI[per=slash]{3e-15}{\meter\per\second\squared} acting on
the test masses. At higher frequencies, the noise associated with the
position measurement limits the sensitivity to about
\SI[per=slash]{10}{\pico\meter\per\rthz}. The sensitivity is further
reduced for frequencies above about \SI{30}{\milli\hertz} by the
transfer function of the detector: the effect of gravitational waves
starts to cancel out as soon as the wavelength of the gravitational
wave approaches an integer multiple of the optical path in the
detector \cite{larson_sensitivity_2000}.

The sensitivity of LISA to changes in the distance between the
spacecraft is low, compared with ground-based
detectors. However, as gravitational waves produce a \emph{strain}, or a
fractional change in distance, the large distance
between the satellites provides a sensitivity to gravitational waves
comparable to those of the much shorter ground based detectors. 

\subsection{Measurement principle}
\label{sec:meas-princ}

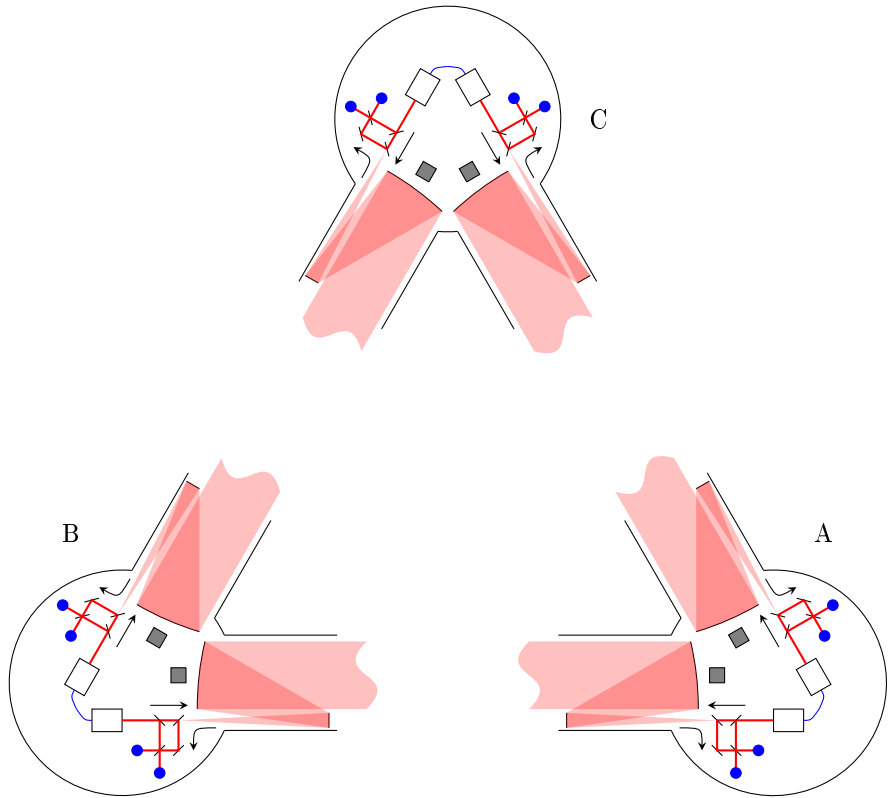
\begin{figure}
  \centering
  \svnidlong 
{$HeadURL: file:///Users/ojennric/SVN/Paper/LISA_Technology/const_scheme.tex $} 
{$LastChangedDate: 2009-06-15 11:26:41 +0200 (Mon, 15 Jun 2009) $} 
{$LastChangedRevision: 50 $} 
{$LastChangedBy: ojennric $} 
\svnid{$Id: const_scheme.tex 50 2009-06-15 09:26:41Z ojennric $} 

\begin{tikzpicture}



\node (A) at (5,-0.5) {A};
\node (B) at (-5,-0.5) {B};
\node (C) at (2,5) {C};

\newcommand{\scfull}{ 
\sctwotelescopes[%
proofmass%
,photodiodes%
,dirarrows%
,laserbackside%
,laserlight%
,interference%
]{}
}

\begin{scope}[shift={(0,5)},rotate=-120,scale=0.5]
  \scfull
\end{scope}

\begin{scope}[rotate around={120:(0,0)}]
  \begin{scope}[shift={(0,5)},rotate=-120,scale=0.5]
    \scfull
  \end{scope}
\end{scope}

\begin{scope}[rotate around={-120:(0,0)}]
  \begin{scope}[shift={(0,5)},rotate=-120,scale=0.5]
    \scfull
  \end{scope}
\end{scope}

\end{tikzpicture}

  \caption{Schematics of the laser links in the LISA constellation. Each satellite
    transmits laser light to the two other satellites and receives
    laser light from the two other satellites. The laser used for
    transmitting is phase-locked to the received laser light,
    establishing a transponder scheme. As transmitted and received
    light share the same telescope, a polarisation multiplexing scheme
    is used to separate the two beams. The use of linear
     polarisations requires slightly different optical benches on the
     sending and the receiving spacecraft. The blue (solid) dots
     indicate where the interferometric measurements are taken. Proof
     masses are shown in grey; the interferometry to determine their
     position has been left out for clarity.}
  \label{fig:LISA_const_scheme}
\end{figure}

\begin{figure}
  \centering
  \svnidlong 
{$HeadURL: file:///Users/ojennric/SVN/Paper/LISA_Technology/part_measurement.tex $} 
{$LastChangedDate: 2009-06-15 11:26:41 +0200 (Mon, 15 Jun 2009) $} 
{$LastChangedRevision: 50 $} 
{$LastChangedBy: ojennric $} 
\svnid{$Id: part_measurement.tex 50 2009-06-15 09:26:41Z ojennric $} 

\begin{tikzpicture}

\newcommand{\schalf}{
    \begin{scope}[scale=0.5]

      \draw (170:3) arc (170:335:3);
      

      \telescope
      \laserlight
      \interference
      \photodiodes
      \PMinterferometry

      \DirectionArrows
      \proofmass

      \draw [point] (bs |- 0,-3) node (SC1A) {};
      \draw [point] (SC1A) ++(0,-2) node (SC1B) {};

      \draw [point] (pm_bs |- 0,1.5) node (SC1C) {};
      \draw [point] (SC1C) ++(0,1.0) node (SC1D) {};

      \draw [point] (PM_center |- 0,1.5) node (PM1A) {};
      \draw [point] (PM1A) ++(0,1.0) node (PM1B) {};

  \end{scope}
}

\schalf
\draw  [dashed] (SC1A) -- (SC1B);
\draw  [dashed] (SC1C) -- (SC1D);
\draw  [dashed] (PM1A) -- (PM1B);

\node [point] (D_SC1) at (SC1B) {};

\draw [->] (SC1D) ++(-0.5,0) -- (SC1D);
\draw [->] (PM1B) ++(0.5,0) -- (PM1B);
\node [anchor=south] at ($ (SC1D)!.5!(PM1B) $)
{\footnotesize{Measurement S/C to test mass}};

\begin{scope}[shift={(8.66,0)},xscale=-1]
  \schalf
  \draw  [dashed] (SC1A) -- (SC1B);
  \draw  [dashed] (SC1C) -- (SC1D);
  \draw  [dashed] (PM1A) -- (PM1B);
  \node [point] (D_SC2) at (SC1B) {};
\draw [->] (SC1D) ++(-0.5,0) -- (SC1D);
\draw [->] (PM1B) ++(0.5,0) -- (PM1B);
\node [anchor=south] at ($ (SC1D)!.5!(PM1B) $)
{\footnotesize{Measurement S/C to test mass}};
\end{scope}

\draw [<->] (D_SC1) -- (D_SC2) node [midway, fill=white]
{\footnotesize{S/C to S/C measurement}};

\end{tikzpicture}

  \caption{Partition of the LISA measurement. Each measurement between
    two test masses is broken up into three different measurements: two
    between the respective test mass and the spacecraft and one
    between the two spacecraft. As the noise in the measurement is
    dominated by the shot noise in the S/C-S/C measurement, the
    noise penalty for the partitioning of the measurement is
    negligible. The blue (solid) dots
     indicate where the interferometric measurements are taken.}
  \label{fig:LISA_partition}
\end{figure}
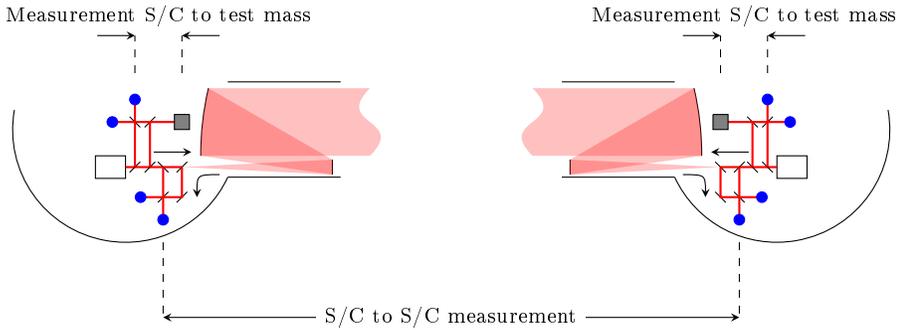
For practical reasons, the interferometric measurement in LISA is
broken up into three distinct parts: the measurement between the
spacecraft, \ie\ between the optical benches that are fixed to the
spacecraft, and the measurement between each of the test masses and
its respective optical bench (see \fref{fig:LISA_partition}). Such a
partition of the measurement would be normally avoided, as it
multiplies the detection noise. However, as the noise budget for LISA
is dominated by the contribution of the shot noise in the measurement
between the spacecraft, this technique only contributes a neglible
amount of noise.

The distance between the test mass and the optical bench is measured
by reflecting light off the proof mass and combining this measurement
beam with a local oscillator on the optical bench (``test mass
interferometer''). 

To measure the distance between the spacecraft, about \SI{2}{\watt} of
infrared light (\SI{1064}{\nano\meter}) is sent through a
\SI{40}{\centi\meter} telescope, used for receiving and transmitting,
to the respective far spacecraft, \ie\ spacecraft A transmits to
spacecraft B and C (nomenclature as in \fref{fig:LISA_const_scheme})
\textit{etc.}, forming six single laser links.  In an ordinary
interferometer, the received light would be directly reflected back to
the transmitting spacecraft A where the light would be combined with a
local oscillator, completing the measurement. However, due to the
large distance between the two spacecraft, a direct reflection of the
light is not feasible. Diffraction widens the transmitted laser beam
to many kilometres at the receiving spacecraft and the received power
is of the order of \SI{100}{\pico\watt} (or about \SI{5e8} photons per
second), resulting in a dilution factor of \num{e-10}.

Therefore, a transponder scheme with offset phase-locking is
implemented.  The received light on spacecraft B is combined with a
local oscillator derived from the transmitting laser (B) and the phase
difference is measured. The frequency of the laser (B) can then be
controlled so that the beat note corresponds to an electronic offset
frequency. This way, the phase of the transmitting laser can be made
to be a true copy of the phase of the received light.  Recombining the
light on spacecraft A then yields a beat note that contains the
electronic offset frequency, the Doppler shift (that can be as large
as \SI[per=slash]{15}{\meter\per\second}) and the signal due to the
gravitational wave  (``science interferometer''). To be able to distinguish a gravitational wave
signal from a noise in the electronic offset frequency, the latter
needs to be  known precisely enough. This is achieved by multiplying the
clock frequency on each spacecraft up to a few \si{\giga\hertz} and
transfering them via sidebands to the other spacecraft where the
sidebands can be compared to the local clock, thus allowing
to assess the clock noise and to correct for it.

Extending the phase-locking scheme to include a phase-locked loop (PLL) between the two
lasers on one spacecraft, it is obvious that all six lasers can be
phase-locked to one (arbitrarily chosen) master laser. Such a scheme
requires of 9 individual phase measurements, two for each arm and one
each between the two lasers on a single spacecraft. In addition each
measurement of the test masses with respect to the optical bench
requires another 6 phase measurements, so that a total of 15 phase
measurements carry the complete information on the gravitational
waves.

\begin{figure}
  \centering
  \setlength{\fboxsep}{0pt}
  
  \begin{minipage}[t]{0.49\linewidth}
    {
      \footnotesize
      \setkeys{Gin}{width=2.5in}
        \input{Fdl}%
    }
  \end{minipage}
  \begin{minipage}[t]{0.49\linewidth}
    \footnotesize
    \setkeys{Gin}{width=2.5in}
      \input{Fdv}
  \end{minipage}
  \par
  \begin{minipage}[t]{0.49\linewidth}
    {
      \footnotesize
      \setkeys{Gin}{width=2.5in}
        \input{Fdalpha}%
    }
  \end{minipage}
  \begin{minipage}[t]{0.49\linewidth}
    \footnotesize
    \setkeys{Gin}{width=2.5in}
      \input{Fpaa}
  \end{minipage}

  \caption{Time evolution of the LISA constellation for a five year
    mission duration. Upper left: Time variation $\Delta l$ of the
    armlengths of the constellation. The variation is limited to about
    $\pm\SI{1}{\percent}$ of the armlength, \ie\
    $\pm\SI{50e6}{m}$. Upper right: relative velocities $\Delta v$
    between the spacecraft in line-of-sight, limited to
    $\pm\SI[per=slash]{15}{\meter\per\second}$. For the chosen laser
    wavelength (\SI{1064}{\nano\meter}),
    \SI[per=slash]{1}{\meter\per\second} corresponds to
    \SI{1}{\mega\hertz} Doppler shift. Lower left: Variation of the
    inner angles of the constellation around the nominal value of
    \ang{60}, limited to $\pm\ang{0.8}$. Lower right: Point-ahead
    angle for the three spacecraft for a nominal orbit. The
    out-of-plane angle $\gamma^\bot$ shows a negligible offset with a
    variation of $\pm\SI{6}{\micro\rad}$ that needs to be corrected
    for the outgoing beam, whereas the in-plane angle $\gamma^\pll$
    displays a large constant offset of about \SI{3.32}{\micro\rad}
    and a small and negligible variation of about \SI{100}{\nano\rad}.}
\label{fig:const_char}
\end{figure}


\svnidlong 
{$HeadURL: file:///Users/ojennric/SVN/Paper/LISA_Technology/Optical.tex $} 
{$LastChangedDate: 2009-06-15 11:26:41 +0200 (Mon, 15 Jun 2009) $} 
{$LastChangedRevision: 50 $} 
{$LastChangedBy: ojennric $} 
\svnid{$Id: Optical.tex 50 2009-06-15 09:26:41Z ojennric $} 

\section{Optical system}
\label{sec:optical-system}

\begin{figure}
  \centering
  \begin{minipage}{0.45\linewidth}
    \includegraphics{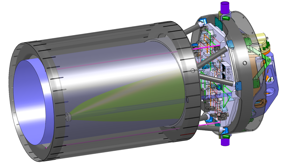}
  \end{minipage}
  \hfill
  \begin{minipage}{0.45\linewidth}
    \includegraphics{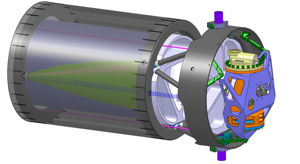}
  \end{minipage}
  \vskip2\baselineskip
  \begin{minipage}{0.45\linewidth}
    \includegraphics{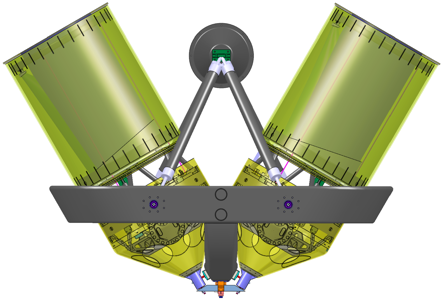}
  \end{minipage}
  \hfill
  \begin{minipage}{0.45\linewidth}
    \includegraphics{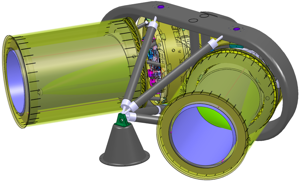}
  \end{minipage}

  \caption{Optical assembly. Telescope with optical bench attached
    (upper row). The optical bench is orthogonal to the telescope axis
    with the optical components facing the main mirror of the
    telescope (upper left). The inertial sensor is attached behind the
    optical bench (upper right) to a support ring also holding the
    optical bench.  A support structure takes the two sub-units (lower row), forming
    the optical assembly. The angle between the two telescopes is nominally 60
    degrees, but can be adjusted within $\pm\SI{1.5}{\degree}$.}
  \label{fig:osa}
\end{figure}

The optical system of LISA contains all the optical components and
their support that are needed for the interferometry. More
specifically, each spacecraft comprises one \emph{optical assembly}
(\fref{fig:osa}, lower row) that consists of two units each
(\fref{fig:osa}, upper row), each made up from the \emph{optical bench},
the \emph{telescope}, and the \emph{gravitational reference sensor} as
well as the associated mounting structures.

The optical bench is mounted parallel to the primary mirror of the
telescope, requiring a non-planar beam path, where the light from the
optical bench to the telescope has to be directed ``up'' to the telescope. 

The Gravitational Reference Sensor (GRS) is mounted behind the optical
bench so that the light from the optical bench to
the GRS has to pass through the optical bench (``down''), resulting in
a non-planar beam path as well.

\subsection{Optical bench}
\label{sec:optical-bench}

\begin{figure}
  \centering
  \svnidlong 
{$HeadURL: file:///Users/ojennric/SVN/Paper/LISA_Technology/OB_func.tex $} 
{$LastChangedDate: 2008-12-15 10:29:41 +0100 (Mon, 15 Dec 2008) $} 
{$LastChangedRevision: 26 $} 
{$LastChangedBy: ojennric $} 
\svnid{$Id: OB_func.tex 26 2008-12-15 09:29:41Z ojennric $} 
\tikzset{ laser/.style={ 
    rectangle, 
    minimum size=1cm, 
    very thick, 
    draw=red!50!black!50, 
    top color=white, 
    bottom color=red!50!black!20, 
  }}
\tikzset{IF/.style ={ circle, 
    minimum size=0.5cm, 
    very thick,
    draw=black}
}

\tikzset{mech/.style = {
    rectangle,
    minimum size=0.5cm,
    very thick,
    draw=black
}}

\tikzset{laser1/.style ={ red, very thick}}
\tikzset{laser2/.style ={ blue, dashed, very thick}}
\tikzset{outport/.style={
    isosceles triangle,
    minimum size =0.5cm,
    very thick,
    draw=black
}}


\begin{tikzpicture}[every label/.style={font=\footnotesize}]

     \coordinate (End1) at ($ (0,1.5) + (11,0) $);
     \coordinate (End2) at ($ (0,0) + (11,0) $);

     \path 
           node (L1) at (0,1.5) [outport]{}
            node [below right= 0.2cm and -0.3cm of L1.west, text
            width=3cm,key] at (L1.south){primary
            laser}
           node (L2) at (0,0) [outport]{}
           node [below right= 0.2cm and -0.3cm of L2.west, text width=3cm,key] at (L2.south) {secondary Laser, from other optical bench}
           node (PW) at (2,3.5) [outport,shape border rotate=90]{}
           node [above,text width=2cm, text centered, key] at ($ (0,4.2) + (0,0)!(PW)!(11,0) $) {Power monitor}
           node (PAAM) at (4,.75) [IF, key, label=right:PAAM]{}
           node (PM) at (7,.75) [IF, key, label=right:TM] {}
           node (Ref) at (9,.75) [IF, key, label=right:Reference] {}
           node (OT) at (5.5,-1) [outport,shape border rotate=-90]{}
           node [below,text width=2cm, text centered, key] at (OT.south){optical truss}
           node (OB) at (6.5,3.5) [outport,shape border rotate=90]{}
           node [above,text width=2cm, text centered, key] at ($ (0,4.2) + (0,0)!(OB)!(11,0) $){other optical bench}
           node (TX) at (4,3.5) [outport,shape border rotate=90]{}
           node [above,text width=2cm, text centered, key] at ($ (0,4.2) + (0,0)!(TX)!(11,0) $) {transmitted beam}
           node (RX) at (9,3.8) [outport,shape border rotate=-90]{}
           node [above,text width=2cm, text centered, key] at ($ (0,4.2) + (0,0)!(RX)!(11,0) $) {received beam}
           node (Sc) [IF,key,label=right:Science] at ($(0,2.5) + (0,0)!(RX)!(11,0) $)  {}
           node (PAAMm) at  ($ (0,2.5) + (0,0)!(TX)!(11,0) $)  [mech] {}
           node [label=right:PAAM, key] at (PAAMm.east) {}; 

      \path node at (0.5,-3) [outport,key, label={right:In/Out}] {}
            node at (4.5,-3) [IF,key, label=right:Interference]{}
            node at (8.5,-3) [mech,key,label=right:Mechanism] {};

      \draw [laser1, ->] (L1) --  ($ (L1)!(PW)!(End1) $) --  (PW);

      \draw [laser1, ->] (L1) --  ($ (L1)!(PAAMm)!(End1) $) --  (PAAMm);
      \draw [laser1, ->] (PAAMm) -- (TX);

      \draw [laser1, ->] (L1) --  ($ (L1)!(OB)!(End1) $) -- (OB);

      \draw [laser1, ->] (RX) -- (Sc);
      \draw [laser1, ->] (L1) --  ($ (L1)!(Sc)!(End1) $) -- (Sc);
      
      \draw [laser1, ->] (L1) -- ($ (L1)!(Ref)!(End1) $) -- (Ref);
      \draw [laser1, ->] (L1) -- ($ (L1)!(PM)!(End1) $) -- (PM);
      \draw [laser1, ->] (L1) -- ($ (L1)!(PAAM)!(End1) $) -- (PAAM);

      \draw [laser2, ->] (L2) -- ($ (L2)!(Ref)!(End2) $) -- (Ref);
      \draw [laser2, ->] (L2) -- ($ (L2)!(PM)!(End2) $) -- (PM);
      \draw [laser2, ->] (L2) -- ($ (L2)!(PAAM)!(End2) $) -- (PAAM);
      \draw [laser2, ->] (L2) --  ($ (L2)!(OT)!(End2) $) -- (OT);           

\end{tikzpicture}

  \caption{Functional diagram of the optical bench. Light from the
    primary laser (red, solid line) is directed to the power monitor, transmitted to the
    telescope via the point-ahead mechanism (PAAM), and is used for interferometric
    measurements of the PAAM, the test mass
    (TM) and the reference interferometer as well as for interfering
    with the received light (Science). The secondary laser (blue,
    dashed line) provides
    light to the optical truss and the interferometric measurements.}

\label{fig:OB_func}
\end{figure}
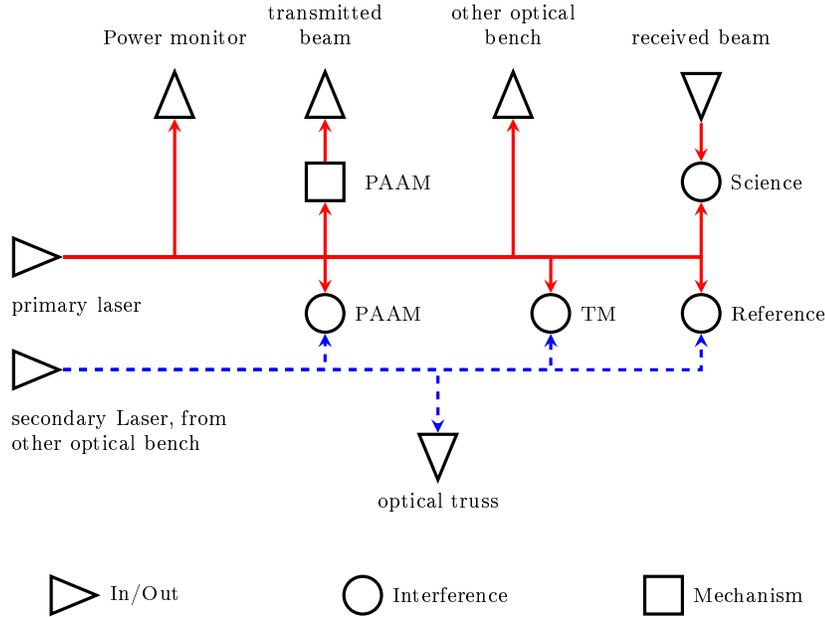

\begin{figure}
  \centering
  \input{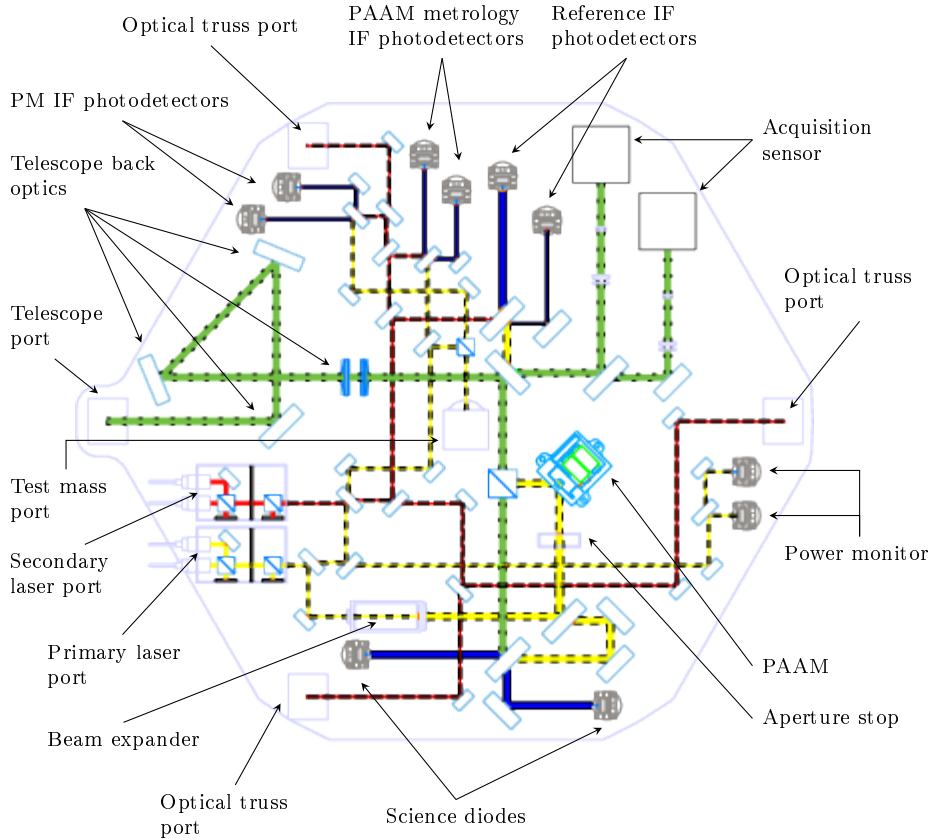}
  \caption{Present baseline architecture for the optical bench. The
    light received from the telescope is depicted green (short dash),
    the light from the primary and secondary laser is depicted yellow
    (normal dash) and red (long dash), respectively. Interfering beams
    are depicted blue (solid line). The optical bench carries the
    telescope back optics, acquisition sensors, the point-ahead
    mechanism (PAAM) and all the optics and detectors needed for the
    interferometry. Beams to the test mass, the optical truss and the
    telescope are orthogonal to the optical bench and exit and enter
    through the respective ports. The width of the beams indicates the
  physical beam diameter.}
  \label{fig:ob}
\end{figure}

\begin{figure}
  \centering
  \begin{minipage}{.49\linewidth}
    \includegraphics[angle=270,width=\linewidth]{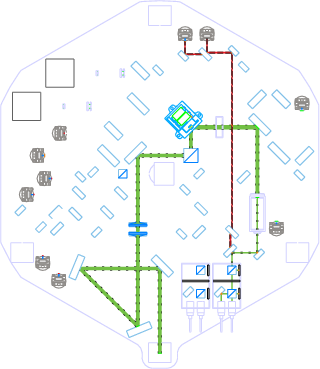}
  \end{minipage}
  \hfill
  \begin{minipage}{.49\linewidth}
    \includegraphics[angle=270,width=\linewidth]{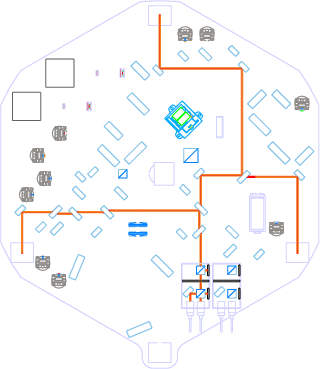}
  \end{minipage}
  \caption{Non-interferometric beam paths on the optical bench. The
    drawing  on the left shows the outgoing beam (green, short dash) and
    the light directed to the power monitors (red, long dash)
    coming from the primary laser. The
    drawing schematics on the right shows the beams coming from the
    secondary laser going to the optical truss ports.}
  \label{fig:NonIF_OB}
\end{figure}

\begin{figure}
\begin{minipage}{.48\linewidth}
    \includegraphics[angle=270,width=\linewidth]{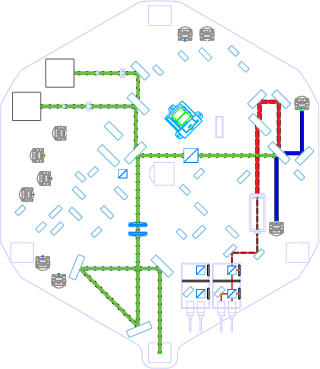}
    \end{minipage}
  \hfill
  \begin{minipage}{.48\linewidth}
    \includegraphics[angle=270,width=\linewidth]{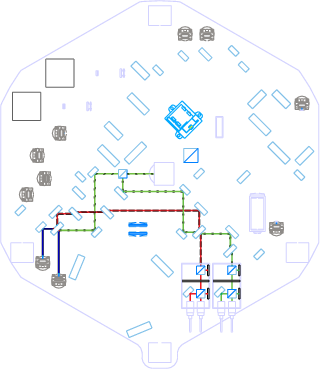}
  \end{minipage}
  \vskip\baselineskip
  \begin{minipage}{.48\linewidth}
    \includegraphics[angle=270,width=\linewidth]{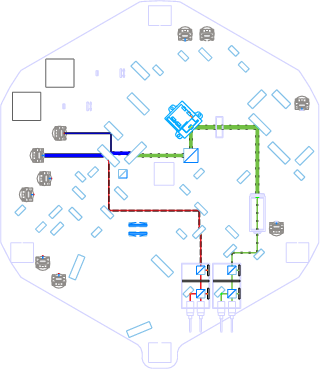}
    \end{minipage}
  \hfill
  \begin{minipage}{.48\linewidth}
    \includegraphics[angle=270,width=\linewidth]{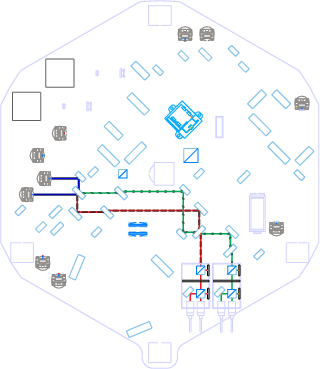}
  \end{minipage}
  \caption{Interferometric beam paths on the optical bench. Reference
    and measurement beams are always shown in green (short dash) and
    red (long dash), respectively. Beam paths after the recombining
    beamsplitter are depicted in blue (solid line). Upper left, the
    main science interferometer is shown with light going to the
    acquisition sensors positioned in the upper right of the optical
    bench, the upper right shows the test mass interferometry. On the
    lower left the metrology interferometer for the point-ahead
    mechanism (PAAM) is shown and the lower right depicts the
    reference interferometer.}
  \label{fig:IF_OB}
\end{figure}

The required functionality of the optical bench (\fref{fig:OB_func})
causes a relatively complex optical bench (\fref{fig:ob}).  

The primary laser (\SI{2}{\watt} at end of life, see
\sref{sec:laser-system}) provides the light to be sent to the far
spacecraft, the reference beam for the science interferometer, and the
measurement beam to the local interferometer, the reference
interferometer and the metrology for the point-ahead mechanism (PAAM, see sec.~\ref{sec:point-ahead-angle}). A small portion of the primary
laser's light is directly send to a laser power monitor so that the
output power of the laser can be measured and controlled.

The secondary
(\SI{100}{\milli\watt}) laser serves as reference beam to the
interferometers used in the PAAM metrology, the measaurement of the
position of the test mass, the reference interferometer and the
optical truss. 

Both, primary and secondary laser are delivered to the optical bench
through single-mode, polarisation maintaining optical fibres. The
polarisation is fixed by a polariser on the optical bench.

Each spacecraft carries two optical benches, so instead of providing
two independent lasers per bench, each bench's primary laser serves as
the other bench's secondary laser. For that, a portion of the light of
the primary laser is brought to the other optical bench through a
fibre that connects the two optical benches (``backside fibre link'').

\subsubsection{Interferometric measurement techniques}
\label{sec:interf-meas-techn}

The interferometric measurements in LISA are based on heterodyne
interferometry, where two laser with respective frequencies $\nu_0$ and
$\nu_0+\Delta\nu$ are combined to yield a beat note with the frequency $\Delta
\nu$, the phase of which is then detected. 

Measurements of
longitudinal displacements can be performed by directing the combined
light on a single element photo detector, whereas for the measurements
of angular displacements, \emph{differential wavefront sensing}
\cite{heinzel_interferometry_2003} is needed. Here, the light is
directed on to a quadrant photo detector (QPD) and the differential
phase between the signals from the different quadrants is used
to determine the angle of the wavefront arriving at the
photodiode. In practise, taking the sum of QPD's signal is used to emulate a
single element photo detector so that one interferometer can measure
both angular displacement and longitudinal displacement simultaneously.

All the interferometers on the optical bench are read out by two
photodetectors, one in each output of the interferometer, providing
redundancy and an increase of the signal-to-noise ratio by a factor of
$\sqrt2$ in the nominal case (\ie\ with both detectors working).

\subsubsection{Science Interferometer}
\label{sec:science-interf}

The science interferometer measures the distance between the receiving
and the transmitting optical bench on two spacecraft.

Most of the primary laser's light passes through a beam expander
necessary to match the beamsize on the optical bench (\SI{5}{mm}) to
the telescope, followed by an aperture stop and the
point-ahead-mechanism (see~\sref{sec:point-ahead-angle}) before encountering first a polarising
beamsplitter set for reflection, and then the telescope ``back
optics'' (see~\sref{sec:telescope}) before encountering the
telescope and been sent out to the receiving spacecraft.

One of the two benches involved
in transmitting and receiving the light carries additionally a
half-wave plate in the path of the transmitted light, resulting in
orthogonal polarisations for the beams in different directions
(\fref{fig:LISA_const_scheme}), so that transmitted and received beam can be
separated by means of a polarising beamsplitter.

As the primary laser provides  the reference beam for the science
interferometer as well, part of the light is split
off after the beam expander and its polarisation rotated by 90 degrees
through another half-wave plate to match the polarisation of the
reference beam to that of the received beam.

The light received from the far spacecraft
($\approx\SI{100}{\pico\watt}$) enters through the telescope and the
telescope port onto the optical bench and is directed from the telescope
back optics directly to the science interferometer where it is
combined with the reference beam (\fref{fig:IF_OB}, upper left). As
the light passes a half wave plate either on transmission or on
reception (but never on both), the received beam is now polarised
orthogonally to the transmitted beam (see
\fref{fig:LISA_const_scheme}). Therefore, the polarising beamsplitter used to
reflect the transmitted beam to the telescope back optics transmits
the received light.

While the optical benches are designed as identical as possible, the
polarisation multiplexing scheme causes a slight difference, as one of
the benches on a spacecraft will have a half wave-plate in the
transmit/receive path where the other doesn't.


\subsubsection{Point-ahead angle mechanism}
\label{sec:point-ahead-angle}

\begin{table}
\caption{Requirements on the point-ahead mechanism (PAAM).}
  \label{tab:paam_req}
  \begin{indented}
    \item[]\begin{tabularx}{\linewidth}{@{}>{\setlength\hsize{.5\hsize}}X>{\setlength\hsize{1.5\hsize}}X}
        \br
        &Requirement\\
        \mr
     Dynamical range&$\pm\SI{700}{\micro\rad}$\\
     Optical path-length stability& $\displaystyle \SI[per=frac]{3}{\pico\meter\per\rthz}\times\uff$\\
     Angular stability&$\displaystyle \SI[per=frac]{16}{\nano\radian\per\rthz}\times\uff$\\
        \br
    \end{tabularx}
  \end{indented}  
\end{table}

The point ahead mechanism (PAAM) compensates the aberration for the beam
transmitted to the far spacecraft (see~\fref{fig:NonIF_OB}). The
aberration, the angle between the received and transmitted light, is
due to the fact that the velocity of the far spacecraft has a
component perpendicular to the line of sight, \ie\ the far
spacecraft appears to move sideways. This component
changes amplitude and direction over the course of time, causing a
time varying aberration. Any constant
aberration could be taken out by a pre-launch alignment process,
however, the variation in the aberration makes an on-orbit mechanism
necessary as a varying angle between the measurement beam and the
local oscillator degrades the contrast on the photo detectors.

The aberration can be decomposed into two components, projected into
the plane of the constellation and perpendicular to that. Most of the
constant aberration of about \SI{3.32}{\micro\radian} is in the plane
of the constellation, most of the variation (but very little offset)
is perpendicular to the plane of constellation and amounts to about
$\pm\SI{6}{\micro\radian}$ (\fref{fig:const_char}). The in-plane offset is
taken care of by pre-launch alignment and the in-plane variation is
small enough to not cause discernible loss of contrast. The out-of-plane component, however,
requires on-orbit correction. It is worthwhile noting that due to the
fact that the optical bench itself is perpendicular to the plane of
the constellation, the out-of-plane angle $\gamma^\bot$ actually lies
in the plane of the optical bench, so that the correction keeps the optical
path planar.

As the aberration depends only on
the orbital dynamics of the constellation, it can be predicted with
high accuracy and  the necessary commands for the point-ahead
angle mechanism can be uploaded to the spacecraft well in advance, so
that no closed-loop controller is foreseen. However, should it be
necessary, such a closed loop controller can be implemented by
evaluating the wavefront at the receiving spacecraft and the
orientation of both sending and receiving spacecraft.

Situated on the optical bench, the PAAM is
exposed to the full magnification of the telescope ($\times 80$), so
that the range of the point-ahead mechanism needs to be
$\pm\SI{480}{\micro\radian}$ for the nominal case,
$\pm\SI{700}{\micro\radian}$, including margin. Furthermore, the
point-ahead mechanism is unavoidably in the optical path of the
science interferometer, so it has to fulfil stringent requirements on
the path-length noise it introduces; the optical path-length noise has
to be smaller than \SI[per=slash]{3}{\pico\meter\per\rthz} for
frequencies above \SI{3}{\milli\hertz}. The angular jitter noise of
the mechanism is required to be smaller than
\SI[per=slash]{16}{\nano\radian\per\rthz} for frequencies above
\SI{3}{\milli\hertz}. Here, the telescope magnification works in
favour of a less stringent requirement, as the angular jitter of the
transmitted beam is reduced by the telescope magnification.

The point-ahead angle and the optical path-length error are monitored
through dedicated metrology on the optical bench (\fref{fig:IF_OB},
lower left), allowing to implement a local control
of the PAAM. 

The currently developed solutions for the PAAM both depend on
monolithic designs (Haberland hinge \cite{haberland_flexure_1981} and bending flexures,
respectively) that have no freely rotating axis but translate elastic
deformation of thin blades into a rotation. This allows to achieve
very little friction and hysteresis for the rotation while keeping a
high stiffness in the other degrees of freedom.

\subsubsection{Test Mass Interferometer}
\label{sec:local-interferometry}

The test mass interferometer  is used to
determine the position of the test mass of the gravitational
reference sensor with respect to the optical bench. Together with the
science interferometer, the test mass interferometer on the
transmitting and the receiving optical bench can be combined to
perform the measurement of the distance between the free-falling proof
masses, \ie\ the test mass interferometer provide an integral part of the
science measurement.

The reference beam
for this interferometer is taken from the secondary laser, the
measurement beam comes from the primary laser. The measurement beam is
directed through the optical bench to the test mass situated in the
gravitational reference sensor at the backside of the optical
bench. The test mass acts as a mirror and reflects the measurement
beam back the optical bench. Similar to the science interferometer,
the measurement beam is out of the plane of the optical bench
(\fref{fig:IF_OB}, upper right).

Separation of the beam going to the test mass and the reflected beam
is again done through polarisation, which is different from the setup
used in LTP that relies on a geometrical separation of the two beams
\cite{heinzel_interferometry_2003}. In LISA, an additional
quarter-wave plate is therefore needed that is passed twice by the
light, effectively resulting in a rotation of the polarisation by
\ang{90}. Another half-wave plate rotates the polarisation of the
measurement beam back by \ang{90} to coincide with the polarisation of
the reference beam and allow interference.

The test mass interferometer provides information on both longitudinal
movement of the test mass with respect to the optical bench (and hence
the spacecraft) and rotation through differential wavefront
sensing. The readout of the test mass interferometer will therefor be
used to feed into the control law of the disturbance reduction system,
augmenting the signals from the capacitive readout (see \sref{sec:dist-reduct-syst-1})

\subsubsection{Optical truss}
\label{sec:optical-truss}

The optical truss interferometry is a method to assess the stability
of the telescope structure (see \sref{sec:telescope}) on orbit. It
consists of three pick-off mirrors separated by \ang{120} on the
mounting structure for the secondary mirror of the telescope, that use
a sample of the outgoing light for a measurement beam. Beamsplitter
and photo detector are co-located with the sampling mirror, the
reference beam is taken from the secondary laser (\fref{fig:NonIF_OB},
right). Taking the measurements at three points allows to reconstruct
the shape of the wavefront of the outgoing light as it is needed for
diagnosis and correction in post-processing.

\subsubsection{Reference Interferometer}
\label{sec:refer-interf}

The reference interferometer provides information on the frequency
noise of the primary laser with respect to the secondary laser. Taking
its measurement beam from the primary laser and its reference beam
from the secondary laser, it is deliberately designed to have unequal
optical path-lengths for the interfering beams so that the relative frequency
noise of primary and secondary laser can be assessed
(\fref{fig:IF_OB}, lower right).

The output signal of the reference interferometer is used to provide
an error signal for the phase-locking of the primary and the secondary
laser.

\subsubsection{Optical bench construction}
\label{sec:optic-bench-constr}

The optical bench is constructed much alike the optical bench for the
LISA Technology Package (LTP) experiment on board LISA Pathfinder
\cite{middleton_prototype_2006,braxmaier_lisa_2004,killow_construction_2006},
with Zerodur as a material for the bench and fused silica for most of
the optical components. Hydroxy-catalysis bonding is used to attach
the optical components to the optical bench. This technology, first
developed for the use in the GP-B mission
\cite{turneaure_development_2003,gwo_ultraprecision_1998}, has found
broad applications as well in ground-based gravitational wave
detectors \cite{amico_fused_2002,smith_mechanical_2003} as \eg\ in
LISA Pathfinder \cite{elliffe_hydroxide-catalysis_2005} due to its
excellent properties regarding dimensional stability of the components
and rigidity and durability of the bond itself.

While Zerodur has the advantage of a very low coefficient of thermal
expansion (CTE) of about $\SI{2e-8}{}$, it is quite brittle and care
has to be taken to restrict the mechanical load on the bench by an
appropriate design of the surrounding structure. The necessary
expertise has been developed during the construction of the optical
bench for LTP.

The few differences between the construction of the optical bench for
LISA with respect to the optical bench for LPF lie in the use of
polarising components for LISA, requiring mounting technology for
different materials, as polarisers are usually not made from fused
silica, and in the inclusion of mechanisms on the optical bench such
as the point-ahead actuator.

\subsection{Telescope}
\label{sec:telescope}

\begin{table}
  \caption{Main requirements on the telescope. The full field of view is
    required for acquisition. The wavefront quality is required only
    for the smaller field of view in the science mode.}
  \label{tab:telescope_req}
  \begin{indented}
    \item[]\begin{tabularx}{\linewidth}{@{}>{\setlength\hsize{.5\hsize}}X>{\setlength\hsize{1.5\hsize}}X}
        \br
        &Requirement\\
        \mr
     Aperture&\SI{40}{\centi\meter}\\[1.5ex]
     Total field of view&\SI{400}{\micro\rad} full angle\\
     \multicolumn{1}{r}{science mode}&$\pm\SI{7}{\micro\rad}$
     out-of-plane\\
     &$\pm\SI{4.2}{\micro\rad}$ in-plane\\[1.5ex]

     Optical pathlength stability& $\displaystyle
     \SI[per=frac]{1}{\pico\meter\per\rthz}\times\uff$\\[1.5ex]
     Magnification&80\\[1.5ex]
     Wavefront quality&$\displaystyle \frac{\lambda}{30}$ for $Z_2$
     and $Z_3$ and the root square sum of $Z_5$ and higher.\\
        \br
    \end{tabularx}
  \end{indented}  
\end{table}

The telescope baselined for LISA is an off-axis telescope with a
\SI{40}{\centi\meter} aperture, a mechanical length of about
\SI{60}{\centi\metre}, and a field of view of $\pm\SI{7}{\micro\rad}$
out-of-plane and $\pm\SI{4}{\micro\rad}$ in plane in which the most
stringent wavefront requirements have to be met. The choice of an
off-axis telescope was driven by the stringent requirement on stray
light that is hard to achieve with a secondary mirror in normal
incidence. Additionally, the off-axis design has the advantage of not
blocking part of the incoming light, thus allowing more light for the
measurement process.

The size of the telescope's aperture is determined by the amount
of laser power required for a given sensitivity, as the size of the
aperture determines both the widening of the beam due to diffraction
and the amount of laser power collected from the received beam. A
diameter of \SI{40}{\centi\meter} results in an equivalent pathlength
noise due to shot-noise of about
\SI[per=slash]{8}{\pico\meter\per\rthz}, the dominating contribution
to the noise budget at frequencies above \SI{3}{\milli\hertz}. 

With a magnification of the telescope of $80$ and a diameter of the
outgoing beam of \SI{400}{\milli\meter}, the input beam to the
telescope has a diameter of \SI{5}{\milli\meter}. A beam expander,
situated on the optical bench (\fref{fig:ob}), matches the typical beam
diameter on the optical bench (\SI{1}{\milli\meter}) to the diameter
required by the telescope.

As equally important as the telescope's ability to gather light is the
quality of the wavefront leaving the telescope.  An ideal, perfectly
spherical wavefront with its centre at the position of the test mass
would render the measurement of the optical pathlength insensitive to
any pointing jitter of the sending spacecraft, as all the radii of a
sphere have the same length. Any deviation from such an ideal
wavefront, however, will translate a spacecraft jitter into an
equivalent pathlength noise.

The point spread function (PSF) that is
customarily used to define the optical quality of a telescope is of
limited use for the LISA telescope. On the one hand, not all the
wavefront distortions that contribute to the PSF are of importance for
LISA, on the other hand, the PSF does not provide enough information
on the distortions that are relevant for LISA. 

The requirements for the wavefront distortion are therefore specified,
somewhat unusual, in terms of the amplitudes of Zernike polynomials with
a root mean square value of less than $\lambda/30$ for $Z_2$ and $Z_3$
each, corresponding to a tilted waveform in the far-field, and a total
of $\lambda/30$ for the root square sum of all other Zernike polynomials
$Z_5$ and higher over the whole field of view.

As the wavefront errors critically depend on the position of the beam
waist with respect to the telescope, the telescope can be refocused on
orbit by adjusting the position of two lenses in the telescope
``back-optics''.

Additionally, the back-optics image the exit pupil of the telescope to
the centre of the test mass, the photodetectors of the science
interferometers and the point-ahead mechanism, minimising the effect
of spacecraft rotation on the science measurement.

An additional complication arises from the fact that the telescopes
for LISA form part of the interferometric path of the science
interferometer, \ie\ any change in optical pathlength between \eg, the
primary and secondary mirror, directly contributes to noise degrading
the science signal.  To reduce the impact of any geometrical
distortions, the optical truss interferometry (see
\sref{sec:optical-truss}) can be used to directly measure the
wavefront of the outgoing beam for later correction in
post-processing.

\subsection{Optical assembly tracking mechanism}
\label{sec:optic-assembly-track}

In addition to the time variation of the angle between the received
and the transmitted light that is caused by the aberration, the angle
between the two telescopes on board one spacecraft changes over time
as well. Nominally \ang{60}, it varies by about \ang{1.5} over the
course of a year due to orbital mechanics (\fref{fig:const_char}). To
compensate for that variation, a mechanism that changes the angle
between the two telescopes is required, the so-called \emph{optical
  assembly tracking mechanism} (OATM), as can be seen in the lower
left of \fref{fig:osa}, connecting the rear ends of the two single
assemblies.

The OATM acts upon the complete assembly of gravitational reference
sensor, optical bench and telescope, rotating the assembly around an
axis perpendicular to the plane of the constellation. This way, the
OATM is \emph{not} part of the optical path of an interferometer,
therefore requiring much less care with regard to introducing
translations in addition to rotations.

The requirements on angular jitter are similar to the residual
spacecraft jitter, \ie\ on the order of a few
\si[per=slash]{\nano\radian\per\rthz}.


\svnidlong 
{$HeadURL: file:///Users/ojennric/SVN/Paper/LISA_Technology/Laser.tex $} 
{$LastChangedDate: 2009-06-15 11:26:41 +0200 (Mon, 15 Jun 2009) $} 
{$LastChangedRevision: 50 $} 
{$LastChangedBy: ojennric $} 
\svnid{$Id: Laser.tex 50 2009-06-15 09:26:41Z ojennric $} 

\section{Laser system}
\label{sec:laser-system}

The laser system currently baselined for LISA  makes
use of the Master Oscillator Fibre Power Amplifier (MOFPA) approach
\cite{zawischa_all-solid-state_1999,weels_narrow-linewidth_2002}. For
LISA, the low power master oscillator is largely identical to the
laser used by the LTP experiment on board LISA pathfinder
\cite{mcnamara_lisa_2008}, a Nd:YAG non-planar ring oscillator (NPRO)
pumped by an internally redundant, fibre-coupled arrangement of laser
diodes. The LTP laser is manufactured by Tesat~GmbH
\cite{bartelt-berger_space_2001}, emitting \SI{30}{\milli\watt} of
\SI{1064}{\nano\meter} light and has been used in a similar
configuration on board the TerraSAR-X satellite
\cite{roth_status_2006,zoran_sodnik_free-space_2006}.

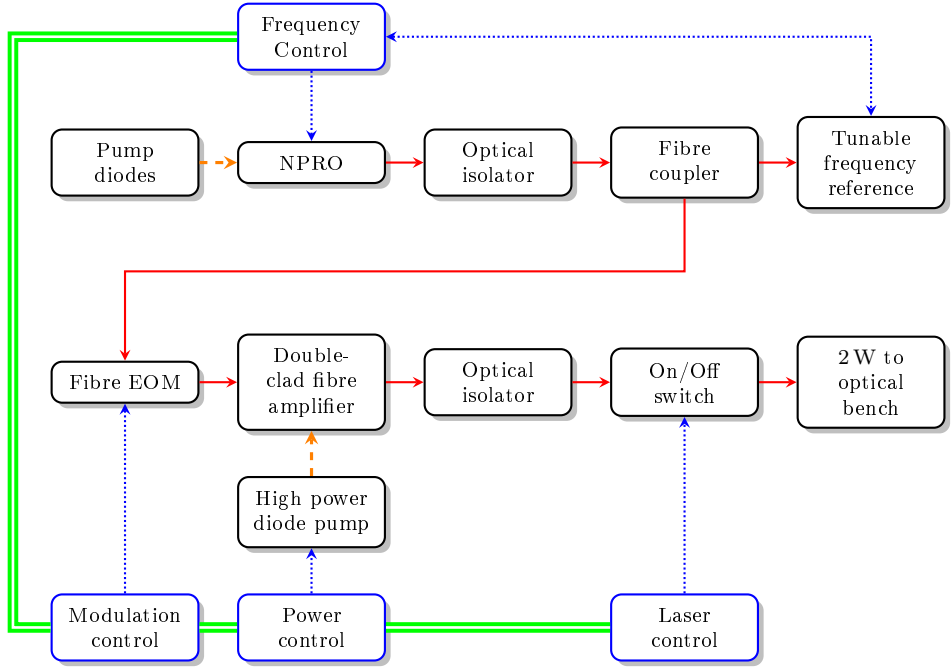
\begin{figure}
  \centering
  \svnidlong 
{$HeadURL: file:///Users/ojennric/SVN/Paper/LISA_Technology/laser_func.tex $} 
{$LastChangedDate: 2009-03-09 17:29:26 +0100 (Mon, 09 Mar 2009) $} 
{$LastChangedRevision: 33 $} 
{$LastChangedBy: ojennric $} 
\svnid{$Id: laser_func.tex 33 2009-03-09 16:29:26Z ojennric $} 

\begin{tikzpicture}[%
  textbox/.style ={rectangle, rounded corners, font=\footnotesize, inner sep=0.5em,
    text width = 1.6cm, text centered, drop shadow, fill=white},%
  tip/.style={->,>=stealth},%
  point/.style={coordinate},%
  box/.style={textbox,thick,draw=black},%
  control/.style={textbox, draw=blue,thick},%
  command/.style={tip, blue, densely dotted, thick},%
  light/.style={tip,red, thick},%
  pump/.style={tip,orange, very thick,dashed},%
  bus/.style={-, green, line width=4pt},%
  every join/.style={rounded corners}%
] 
\matrix[column sep=.5cm, row sep=.3cm] { 
  &&
  \node [point] (p0){};\\
  & & \node[control] (freq_control) {Frequency Control}; &\\
  \\
  &
  \node[box] (pump) {Pump diodes}; &
  \node[box] (npro)  {NPRO}; &
  \node[box] (isolator)  {Optical isolator}; &
  \node[box] (coupler)  {Fibre coupler}; &
  \node[box] (reference) {Tunable frequency reference}; \\[1.5em]
  & & &
  \node [point] (p1) {};\\[1.5em]
  \node[point] (p2) {}; &
  \node[box] (EOM)  {Fibre EOM};&
  \node[box] (amplifier)  {Double-clad fibre amplifier}; &
  \node[box] (isolator2)  {Optical isolator}; &
  \node[box] (switch)  {On/Off switch}; & 
  \node[box] (output)  {\SI{2}{W} to optical bench}; \\
  \\
  &
  &\node[box] (pump2)  {High power diode pump}; \\
  \\
  &
  \node[control] (mod_control)  {Modulation control}; &
  \node[control] (power_control)  {Power control}; & &
  \node[control] (switch_control) {Laser control}; &
  \\
};

 



\draw [pump] (pump)--(npro);
\draw [light] (npro)--(isolator);
\draw [light] (isolator)--(coupler);
\draw [light] (coupler.south) |- (p1) -| (EOM.north);
\draw [light] (EOM)--(amplifier);
\draw [light] (amplifier)--(isolator2);
\draw [light] (isolator2)--(switch);
\draw [light] (switch) -- (output);

\draw [pump] (pump2) -- (amplifier);
\draw [light] (coupler) -- (reference);

\draw [command] (power_control) -- (pump2);
\draw [command] (switch_control) -- (switch);
\draw [command] (mod_control) -- (EOM);
\draw [command] (freq_control.south) -- (npro.north);
\draw [command,<->] (freq_control.east) -| (reference.north);

\draw [bus] (freq_control.west) -| (p2) |- (mod_control) --
(power_control) -- (switch_control);

\draw [-, white, line width=1pt] (freq_control.west) -| (p2) |- (mod_control) --
(power_control) -- (switch_control);
-, green, line width=4pt

\end{tikzpicture}

  \caption{Schematics of the LISA laser system. Laser light is denoted
    by the red lines (solid), starting from the NPRO laser via the
    electro-optic modulator (EOM) to the
    fibre amplifier on to the optical bench. The pump light is shown
    orange (dashed). Control inputs are denoted with blue lines
    (dotted), the spacecraft bus connecting the control units in green
    (double line).}
  \label{fig:laser_system}
\end{figure}

The light of the NPRO passes an optical isolator to suppress optical
feedback  and is coupled into two optical single-mode
fibres, the smaller fraction of the light is taken to be used for
prestabilisation purposes, the larger fraction fed into a fibre-based electro-optical modulator,
that imprints sidebands used for clock-transfer and ranging. From
there, it enters a double-clad fibre amplifier, pumped by a redundant
array of fibre-coupled laser diodes, bringing the laser power up to
the required \SI{2}{\watt}.  After passing another optical isolator
and an on/off switch controlled by the spacecraft computer, the light
is then delivered via an optical fibre directly to the optical bench
(\fref{fig:laser_system})

As polarisation encoding is used in LISA to distinguish transmitted
from received light, the light entering the optical bench needs to be
linearly polarised. This is ensured by a polariser as the first component on
the optical bench. To avoid unnecessary stray-light, \SI{98}{\percent}
of the light power arriving on the optical bench needs to be in the linear
polarisation transmitted by the polariser.

As the position of the test masses is read out interferometrically in
the test mass interferometry, the requirement on the permissible
acceleration noise for the test masses leads to a requirement on the
power stability of the laser, as a variation in laser power causes a
variation in radiation pressure on the test masses and therefore a
variation in the acceleration.  Consequently, the intensity noise for
the laser has to be smaller than
$\SI[per=slash]{1}{\milli\watt\per\rthz}$ at \SI{3}{\milli\hertz}.  A
summary of the requirements on the laser at end-of-life is given in
\tref{tab:laser_req}.

\begin{table}
\caption{Laser requirements, specified at end-of-life.}
  \label{tab:laser_req}
  \begin{indented}
    \item[]
      \begin{tabularx}{\linewidth}{@{}>{\setlength\hsize{.5\hsize}}X>{\setlength\hsize{1.5\hsize}}X}
        \br
        &Requirement\\
        \mr
        Wavelength& \SI{1064.5}{\nano\meter}\\[1.5ex]
        Output power                 & \SI{2}{\watt} \\
        Polarisation             & linear, containing more than \SI{98}{\percent} of
        optical power in main polarisation                                       \\
        Power stability & $\displaystyle \SI[per=frac]{1}{\milli\watt\per\rthz}\times\uff$ \\
        \br
    \end{tabularx}
  \end{indented}  
\end{table}

\subsection{Laser Frequency noise suppression}
\label{sec:laser-freq-noise}

Frequency stabilisation of the lasers is a vital part of the LISA measurement
scheme. As in any interferometric length measurement, a frequency noise $\delta
\nu$ causes an equivalent noise in the length measurement $\delta x$
that is proportional to the difference of optical pathlength $\Delta
L$ and the fractional frequency noise.
\begin{equation}
  \label{eq:1}
  \delta x = \Delta L \frac{\delta\nu}{\nu}
\end{equation}

The difference in optical pathlength in LISA can be as large as
$\Delta L=\SI{e8}{m}$ due to the orbital motion of the spacecraft and
the equivalent pathlength noise contribution allocated to frequency
noise is around $\delta\tilde x
=\SI[per=slash]{0.4}{\pico\metre\per\rthz}$ at \SI{3}{mHz}. This
results in a required frequency stability of $\delta\tilde\nu =
\SI[per=slash]{1.2e-6}{\hertz\per\rthz}$ at \SI{3}{mHz}. Starting from
a free-running laser that has a typical frequency noise of
$\delta\tilde\nu_\text{free}=\SI[per=slash]{3}{\mega\hertz\per\rthz}$
at \SI{3}{mHz}, such a reduction of frequency noise by about 12 orders
of magnitude is difficult to achieve in a single step.

Therefore, a three level approach has been chosen for LISA. First, a
pre-stabilisation of the free-running laser to a level of
\SI[per=slash]{30}{\hertz\per\rthz}, then a stabilisation of the laser
to the LISA arms and finally the post-processing stage as a last step.



\begin{table}
\caption{Frequency stabilisation requirements}
  \label{tab:freq_stab_req}
  \begin{indented}
    \item[]\begin{tabular}{@{}l>{$\displaystyle}r<{$}@{\,}>{$\displaystyle\times\,}l<{$}}
        \br
        Stabilisation stage&\multicolumn{2}{l}{Performance after
          stabilisation in \si[per=frac]{\hertz\per\rthz}}\\
        \mr
        Free running&\frac{\SI{e4}{\hertz}}{f} \\
        Pre-stabilisation&\frac{\SI{30}{\hertz}}{f}&\uff\\
        Arm locking&\SI{.3}{}&\uff\\
        TDI&\SI{4e-7}{}&\uff\\
        \br
    \end{tabular}
  \end{indented}  
\end{table}

\subsubsection{Prestabilisation}
\label{sec:prestabilisation}

The prestabilisation is the first stage of the frequency stabilisation
scheme for LISA.  It requires a local frequency reference, such as a
cavity \cite{livas_tunable_2007,thorpe_laser_tbp,mueller_laser_2005},
a molecular resonance
\cite{leonhardt_space_2006,leonhardt_iodine_2006} or a dedicated
heterodyne interferometer with unequal arms (much like the one
employed in the LISA Pathfinder experiment
\cite{wand_noise_2006,heinzel_interferometry_2006})

Laser stabilisation to a cavity using a variety of techniques, most
prominently RF-sideband locking has been demonstrated to well beyond
the required stability for LISA (see \eg\ \cite{notcutt_simple_2005}
for a demonstration of a stability of
$\delta\tilde\nu=\SI[per=slash]{1}{\hertz\per\rthz}$ for frequencies
above \SI{1}{\hertz}) at frequencies somewhat higher than the LISA
frequency band. In the LISA frequency band, thermally driven changes
of the cavity length are a major contributor to the residual frequency
noise. As the thermal environment for LISA will be exceptionally
stable, this is mainly a problem for laboratory-based demonstration or
verification experiments, as those need sophisticated thermal
insulation to reach the required stability. Using multiple-stage
insulation systems, a frequency stability of $\tilde\nu =
\SI[per=slash]{30}{\hertz\per\rthz}$ at \SI{3}{\milli\hertz} has been
demonstrated at NASA's Goddard Space Flight Center (GSFC)
\cite{mueller_laser_2005}. The cavities used in this experiment
underwent environmental testing and the given performance has been
reached before and after the testing cycle.

The stabilisation of a frequency-doubled Nd:YAG laser to a hyperfine
absorption line of the $\text{I}_2$ molecule has a long history as
well. Typically used for comparing absolute frequencies in metrology,
(\cite{nevsky_frequency_2001,hong_portable_1998}), iodine
stabilisation has been employed in ground-based gravitational wave
detectors \cite{musha_short-_2000} and is currently under
investigation for applications in LISA
\cite{leonhardt_space_2006,leonhardt_iodine_2006,mondin_validation_2004}.
where frequency stability of around $\SI{10}{}\ldots
\SI[per=slash]{100}{\hertz\per\rthz}$ in the frequency range of
$\SI{1}{}\ldots\SI{100}{\milli\hertz}$ has been demonstrated in tabletop
experiments \cite{musha_short-_2000, leonhardt_iodine_2006}.

In contrast to the stabilisation on a cavity resonance, stabilisation
on a molecular line provides an absolute frequency reference; the drawback is
some added complexity due to the need of frequency-doubled
light. Recently, a frequency-doubling system has been qualified for
space application in the framework of the technology
development for the SIM mission \cite{chang_waveguide_2007}, greatly
reducing the impact of frequency doubling on the technology
development for LISA.

Heterodyne interferometry, as opposed to the ``homodyne''
stabilisation schemes described above, does not require a tuning of
the laser to the reference, as it provides an error signal largely independent of
the common-mode frequency of the light used. The drawbacks are the
need for two light fields, separated by the heterodyne frequency and
the comparatively low sensitivity. The use of a heterodyne
interferometer with optical paths deliberately chosen to be unequal
has been proposed for LISA, using a scheme much alike the the
reference interferometer in the LTP experiment on board LPF.

\subsubsection{Arm-locking}
\label{sec:arm-locking}

A possible second stage of the frequency stabilisation scheme uses the
interferometer arms of LISA as a frequency reference. By design, the
fractional stability of the arms in the frequency range of around \SI{1}{mHz} is
on the order of $\delta \tilde x/L \sim
\SI[per=slash]{e-21}{\per\rthz}$ and is guaranteed to fulfil the
science requirements for LISA.

Arm-locking therefore makes use of this stability and derives an error
signal from the phase-difference of the local laser and the received
light. As the received light is phase-locked to the local laser at the
remote spacecraft, it can be regarded to carry a replicate of the
noise of the local laser delayed by one full round-trip time
$\tau=\SI{33}{s}$ \cite{sheard_laser_2003}. After choosing a suitable
control law, the noise is suppressed at frequencies $f$ smaller than
the corresponding round-trip frequency $f_0=1/\tau=\SI{30}{mHz}$ but
causes significant amplification of the noise at integer multiples of
$f_0$ \cite{sylvestre_simulations_2004} as well as a long decay time
for the initial conditions. A more elaborate implementation of
arm-locking \cite{herz_active_2005} uses the phase-differences from
both arms in sum and difference to suppress the noise spiking.

The main advantage of the arm-locking scheme is the additional
suppression of the laser frequency noise without the need of any
additional hardware, as the sensors for the required phase
measurements and the actuators for setting the laser frequency are already
present. The control law is fully implemented in software and requires
no additional resources.

A proof-of-concept implementation in hardware use RF signals instead
of light and a \SI{300}{m} coaxial cable to simulate the LISA arm
\cite{marn_phase_2005} and shows the feasibility of unity gain
frequencies above the inverse of the delay time
($\tau=\SI{1.6}{\micro\second}$ as well as the predicted amplification
of the noise and the ``ringing'' after lock acquisition. Similar
experiment, using light in optical fibres ($L=\SI{10}{\kilo\metre}$,
$\tau=\SI{100}{\micro\second}$) and purely electronical delays
\cite{thorpe_arm-locking_2006} yield comparable results.

\subsubsection{Time delay interferometry}
\label{sec:time-delay-intef}

The third stage of the frequency stabilisation scheme, time-delay
interferometry (TDI), does not reduce the laser frequency noise
\emph{in situ}, but rather suppresses the effects of laser frequency
noise in a post-processing stage. In contrast to standard
interferometers, where the light from the two arms is combined
optically and the phase of the individual light impinging on the
recombining beamsplitter is not known, in LISA each incoming light is
combined optically with a reference beam individually, so that the
phase of the incoming light is individually measured and
recorded. This allows to make use of correlations in the frequency
noise and to remove the frequency noise down to the level of the
measurement accuracy provided for the individual phase measurements by
algebraically combining phase measurements data delayed by multiples
of the distance between the spacecraft to the so called \emph{TDI
  variables}.

It must be pointed out, that the ability to use the individual phase
measurements in post-processing does \emph{not} depend on the actual
values of the measurements. This means that TDI is not in any way
restricted by arm-locking (or does in any way restrict arm-locking,
for that matter) \cite{shaddock_postprocessed_2004}.

The first implementation of the algorithm was based in the frequency
domain and dealt with a much simplified constellation
\cite{giampieri_algorithms_1996}. Such a frequency-domain based
implementation is difficult to generalise to the case of
changing arm-length differences and more complex interferometric schemes.

Subsequent implementations of the algorithm have therefore been based
in the time-domain and include signals from all three spacecraft
\cite{tinto_cancellation_1999,armstrong_time-delay_1999}. The simple
time-domain implementation of the TDI algorithm (``first generation
TDI'') using only phase measurement data delayed by the respective
distances between the spacecraft only cancels the frequency noise
exactly for fixed inter-spacecraft distances (much like the algorithm in
the frequency domain) and requires a initial frequency noise of the
lasers not larger than \SI[per=slash]{5}{\hertz\per\rthz}
\cite{cornish_effects_2003}. Further refinements of the algorithm
(``second generation TDI'') allow to deal with changing arm-lengths as
well \cite{tinto_time-delay_2002,shaddock_data_2003} by using phase
measurements data that are delayed by multiples of the inter-spacecraft
distances.

Using TDI with changing
arm-lengths requires in addition the ability to perform phase measurements at
arbitrary times to accommodate for the fact that the travel time of the
light between the spacecraft will not only be different for each arm, but also
changing over time. This additional complication can be overcome
by oversampling and subsequent high-precision interpolation
\cite{shaddock_postprocessed_2004} of the phase measurements. 

A rigorous algebraic approach to the mathematics of TDI progressed as
well from considering a purely static constellation
\cite{dhurandhar_algebraic_2002} to coping with changing arm-lengths
\cite{nayak_algebraic_2004} and a fully relativistic treatment of the
optical links \cite{dhurandhar_general_2008}. The set of TDI variables
forms a complete set of interferometric observables, so that any
interferometric combination can be retrieved by linearly combining
suitable TDI variables \cite{dhurandhar_algebraic_2002}. Furthermore,
suitably chosen linear combinations of TDI variables correspond to
optimal statistical inference \cite{romano_principal_2006}.

A review of the current
state-of-the art techniques and the mathematical understanding of the
algorithm can be found in \cite{dhurandhar_time-delay_2005}.

\paragraph{Experimental demonstration}
\label{sec:exper-demonstr}

A full experimental demonstration of TDI is difficult to achieve,
mainly because of the need to provide sufficient, or at least
representative, time delays between the data streams. As using a
propagation delay through free space is clearly out of the question,
all experimental approaches use electronic delays of the measured
signal to emulate the optical delay. Using this techniques,
experimental demonstrations of the TDI with a delay of \SI{2}{s} and
later \SI{16}{s} \cite{cruz_lisa_2006,cruz_time_2006a} showed a
reduction of the laser phase noise by 5 orders of magnitude at around
\SI{1}{mHz}, coming close to the required LISA performance.


\svnidlong 
{$HeadURL: file:///Users/ojennric/SVN/Paper/LISA_Technology/Phase.tex $} 
{$LastChangedDate: 2009-06-15 11:26:41 +0200 (Mon, 15 Jun 2009) $} 
{$LastChangedRevision: 50 $} 
{$LastChangedBy: ojennric $} 
\svnid{$Id: Phase.tex 50 2009-06-15 09:26:41Z ojennric $} 

\section{Phase measurement}
\label{sec:phase-measurment}


The phasemeter for LISA is one of the few payload items that can claim
no or very little heritage from LISA Pathfinder. The main reason lies
with the fact that LISA Pathfinder uses a relatively low heterodyne
frequency of about \SI{1.6}{\kilo\hertz}
\cite{heinzel_interferometry_2003,heinzel_ltp_2004} in its
interferometers, the heterodyne frequency for LISA is much higher due
to the relative motion of the spacecraft and the resulting Doppler
effect. The requirement for the LISA phasemeter calls for a maximum
admissible frequency heterodyne frequency of \SI{15}{\mega\hertz} and
for a frequency rate of up to
\SI[per=slash]{1}{\hertz\per\second}. Additionally, the phasemeter
must be compatible with data transfer and ranging tones on the laser
link between the spacecraft as well as with the transmission of the
clock signal, none of which are present on LISA
Pathfinder.


One way to address the challenge of measuring the phase of a
heterodyne signal at around \SI{10}{\mega\hertz} is by a counting and
timing technique, where the integer numbers of cycles in a longer time
period (\eg\ \SI{10}{\milli\second} corresponding to \SI{100}{\hertz}
measurement frequency) are counted and the remaining fraction of a
cycle is measured by a high precision timer. Such as scheme has been
demonstrated
\cite{jennrich_demonstration_2001,pollack_demonstration_2006} and
proven to be compatible with data transfer between spacecraft
\cite{pollack_demonstration_2006-1}.

\begin{figure}
  \centering
   \svnidlong 
{$HeadURL: file:///Users/ojennric/SVN/Paper/LISA_Technology/phasemeter.tex $} 
{$LastChangedDate: 2008-12-15 10:35:02 +0100 (Mon, 15 Dec 2008) $} 
{$LastChangedRevision: 27 $} 
{$LastChangedBy: ojennric $} 
\svnid{$Id: phasemeter.tex 27 2008-12-15 09:35:02Z ojennric $} 

\begin{tikzpicture}[%
  small text/.style={font=\footnotesize},%
  every label/.style={small text},%
  textbox/.style ={rectangle, small text, inner sep=0.25em, text width = 1.6cm, text centered},%
  tip/.style={->,>=stealth},%
  point/.style={coordinate},%
  skip loop/.style={to path={.. ++(.2,0) .. (\tikztotarget)}},%
  box/.style={textbox,thick,draw=black},%
  LPF/.style={textbox,thick,draw=black, text width=0.8cm},%
  RF/.style={circle, thick, draw=black,minimum size=0.6cm},%
  signal/.style={tip, thick,draw=black},%
  divide/.style={-,dashed,thick, draw=black}%
  ]
  
  \matrix[column sep=0.3cm, row sep=0.2cm]{
    &
    & 
     \node [point] (p_analog_north) {};& 
    & 
    & 
    &
    &
    &
    \node [point] (p_fpga_north) {};
    \\

    &
    & 
    & 
    & 
    \node [point] (p0) {}; & 
    \node [RF] (I_LO) {$\sim$} ;\\

    &
    & 
    & 
    & 
    &
    \node [RF] (I_MX) {}; 
    \draw [-,thick] (I_MX.north east)--(I_MX.south west);
    \draw [-,thick] (I_MX.north west)--(I_MX.south east); &
    \node [box,
    label=above:{$\SI{50}{\mega\hertz}\to\SI{10}{\kilo\hertz}$}] 
    (dec_I) {Down sampling};&
    \node [point] (p_I) {};\\[-0.3cm]

    \node [point] (PD) {}; \draw [signal,-] ($(PD)-(0,0.3)$) --
    ($(PD)+(0,0.3)$);
    \draw [signal,-] ($(PD)-(0,0.2)$) arc (-90:90:0.2);&
    \node [LPF] (LPF1) {Anti-alias};&
    &
    \node [box, text width=1cm] (ADC) {\SI{50}{\mega\hertz} ADC};&
    \node [point] (pX) {};&
    \node [point] (p1) {};
    &
    &
    &
    &
    \node [box] (phase) {Phase reconstruction}; &
    \node [box,
    label=above:{$\SI{10}{\kilo\hertz}\to\SI{100}{\hertz}$}] (dec3)
    {Down sampling};&
    \node [point] (out) {};
    \\

    &
    & 
    & 
    & 
    &
    \node [RF] (Q_MX) {}; 
    \draw [-,thick] (Q_MX.north east)--(Q_MX.south west);
    \draw [-,thick] (Q_MX.north west)--(Q_MX.south east); &
    \node [box,
    label=below:{$\SI{50}{\mega\hertz}\to\SI{10}{\kilo\hertz}$}] 
    (dec_Q) {Down sampling};&
    \node [point] (p_Q) {};\\

    &
    & 
    & 
    & 
    \node [point] (p5) {}; & 
    \node [RF] (Q_LO) {$\sim$} ;&
    \\

    &
    &
    &
    &
    &
    &
    \node [box] (update) {Update LO frequency};\\

    &
    &
    &
    &
    &
    &
    \node [point] (p7) {};\\

    &
    &
    \node [point] (p_analog_south) {};&
    &
    &
    &
    &
    &
    \node [point] (p_fpga_south) {};\\
  };

  \draw [signal] ($(PD)+(0.2,0)$) -- (LPF1);
  \draw [signal] (LPF1) -- (ADC);
  \draw [signal] (ADC) -| (I_MX);
  \draw [signal] (ADC) -| (Q_MX);

  \draw [signal] (I_LO) -- (I_MX);
  \draw [signal] (I_MX) -- (dec_I);
  \draw [signal] (dec_I) -- (p_I) |- (phase.165);

  \draw [signal] (Q_LO) -- (Q_MX);
  \draw [signal] (Q_MX) -- (dec_Q);
  \draw [signal] (dec_Q) -- (p_Q) |- (phase.195);

  \draw[signal] (phase) -- (dec3);
  \draw[signal] (dec3) -- (out);

  \draw[signal] (update.south) -- (p7) -| (phase.south);
  \draw[signal] (p_Q) |- (update.east);
  \draw[signal] (update.west) -| (Q_LO.south);

   \draw[signal,-] (update.west) -| ($ (pX) - (0,0.2) $);

   \draw[signal,-]  ($ (pX) - (0,0.2) $) arc (-90:90:2mm);
   \draw[signal] ( $(pX) + (0,.2) $) |- (I_LO);

   \draw[divide] (p_analog_south) -- (p_analog_north);
   \draw[divide] (p_fpga_south) -- (p_fpga_north);

  \node[draw=none, small text, text centered] at ($ (phase.165) - (3em,0)$) {$I$};
  \node[draw=none, small text, text centered] at ($ (phase.195) - (3em,0)$) {$Q$};

  \node[draw=none, small text, right, inner sep =0pt] at ($(PD)+ (0,0.8)$)  {Photoreceiver};

  \node[draw=none, label={right:Local oscillator ($I$)}] at
  ($(I_LO)+(.5em,0)$) {};
  \node[draw=none, label={right:Local oscillator ($Q$)}] at
  ($(Q_LO)+(.5em,0)$) {};

  \node[draw=none, label=left:analogue] at (p_analog_north) {};
  \node[draw=none, label=right:FPGA] at (p_analog_north) {};
  \node[draw=none, label=left:FPGA] at (p_fpga_north) {};
  \node[draw=none, label=right:{floating point processor}] at (p_fpga_north) {};

\end{tikzpicture}

   \caption{Block diagram of the LISA phasemeter. Signals from the
     photodetector pass an analogue anti-alias filter before
     digitisation in an \SI{50}{\mega\hertz} analogue-digital
     converter (ADC) and further processing to determine phase and
     frequency, based on integer arithmetic in a field programmable
     gate array (FPGA). A phase
     reconstruction algorithm to correct residual tracking errors is
     implemented in a floating point processor and feeds back to the
     local oscillator.}
  \label{fig:phasemeter}
\end{figure}
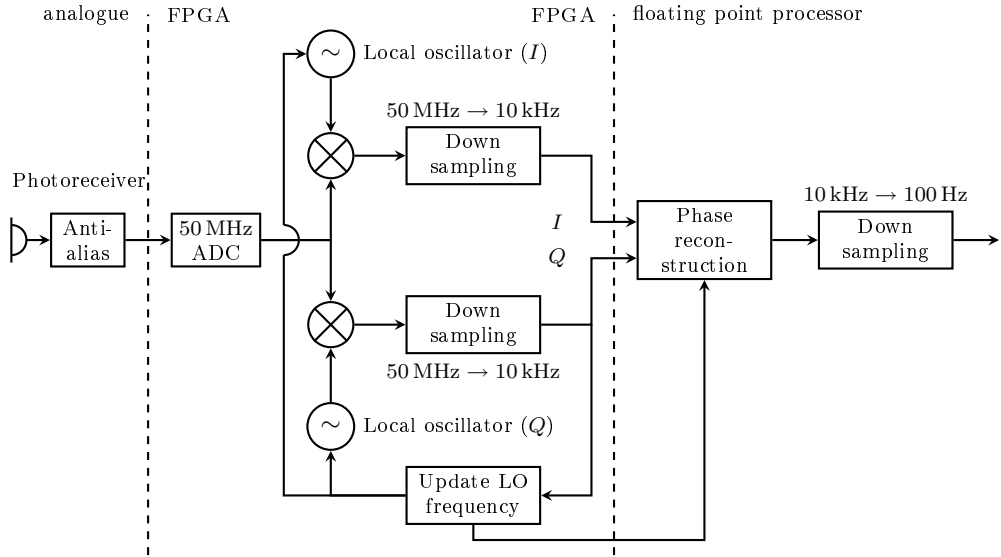

The phasemeter architecture baselined for LISA is based on a more
traditional digital
phase locked loop (DPLL) as sketched in \fref{fig:phasemeter}.  The
signal from the photoreceiver passes through an analogue anti-alias
filter and is then digitised at \SI{50}{\mega\hertz} on an \SI{8}{bit}
analogue-digital-converter (ADC). The digitization frequency has to be
chosen high enough to exceed the Nyquist frequency for the highest
occurring beat note in the system.

The signal is then multiplied with a local oscillator whose
frequency is close to the signal frequency. The down-sampled, \ie\
low-pass filtered and decimated, 
output of this multiplication is directly proportional to the
\emph{phase} difference between signal and local oscillator and is
used as an error signal to drive the frequency and phase of the local
oscillator to be the same as for the signal. The DPLL operates at
\SI{10}{\kilo\hertz}, updating the local oscillator every
\SI{0.1}{\milli\second}, fast enough to follow the frequency changes
occurring in LISA. For performance reasons, these operations are
implemented in a field programmable gate array (FPGA) and all
operations are based on integer arithmetic.

Residual tracking errors are corrected by evaluating the information
in the two quadratures of the error signal in a floating point
processor and combine those with the local oscillator phase. Further
filtering of the signal yields the output at a rate of
\SI{100}{\hertz} for recording.  A more detailed insight into the
principle of operation of the phasemeter including initial results on
simulated data is given in \cite{shaddock_overview_2006}.


\svnidlong 
{$HeadURL: file:///Users/ojennric/SVN/Paper/LISA_Technology/DRS.tex $} 
{$LastChangedDate: 2009-06-15 11:26:41 +0200 (Mon, 15 Jun 2009) $} 
{$LastChangedRevision: 50 $} 
{$LastChangedBy: ojennric $} 
\svnid{$Id: DRS.tex 50 2009-06-15 09:26:41Z ojennric $} 

%

\section{Disturbance Reduction System}
\label{sec:dist-reduct-syst-1}

The disturbance reduction system (DRS) of LISA is one of the main
components of the mission. Whereas the interferometric measurement
system allows to measure the distance between the test masses to
picometer accuracy, the DRS is responsible to render these
measurements meaningful, as it ensures that the test masses follow as
much as possible gravitational orbits, \ie\ experience as small as
possible accelerations.

Thus the DRS consists of the gravitational reference sensor and its
ancillary structures, the micro-Newton propulsion system that is used
to provide the thrust for the fine attitude and position control of
the spacecraft, and the
control law that takes the data from the gravitational reference
sensor and controls the micro-Newton thrusters such as to keep the
spacecraft centred on the test mass.

\subsection{Principle of operation}
\label{sec:principle-operation}

The DRS main objective is to maintain the free fall of a test mass that
serves as nominal reference point for the measurement of the
inter-spacecraft distance. To keep the test mass in free fall, the
DRS measures the position and orientation of the test mass with
respect to the spacecraft, applies a control law and commands
micro-newton thrusters so that the test mass remains in its nominal
position with respect to the spacecraft.

Situated inside the spacecraft, the test mass is shielded from the
external effects, such as solar radiation
pressure and (to a certain degree) the interplanetary magnetic
field. In addition, the spacecraft architecture has to ensure that the
forces on the test mass are as small as possible, requiring special
design precautions regarding the mass distribution, the thermal
balance and the magnetic cleanliness.

\begin{figure}
  \centering
  \input{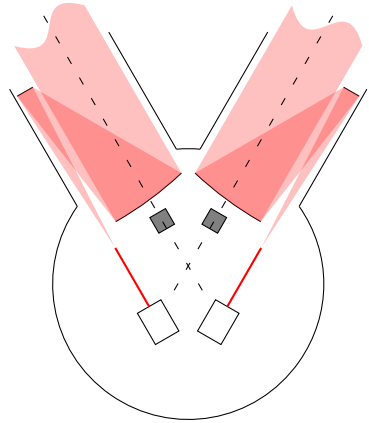}
  \caption{Position and orientation of the two test masses (dark grey)
    in the spacecraft. The sensitive axes of the sensors are indicated
    with the dashed lines and are aligned with the direction to the
    far spacecraft.}
  \label{fig:sensitive_axis}
\end{figure}

Each spacecraft has two gravitational reference sensors (GRS), each
mounted in the line of sight of the corresponding telescope
(see~\fref{fig:osa}, upper right panel), behind the optical bench. The
\emph{sensitive axis} of the DRS denotes the axis aligned to the line of
sight to the telescope and consequently to the test mass in the remote
spacecraft (see~\fref{fig:sensitive_axis}). As LISA employs two test
masses per spacecraft, it is impossible to keep both of them in free
fall condition in all degrees of freedom and ensure at the same time
that the test masses stay close to their nominal position. However, it
is sufficient to maintain free fall in the direction of the sensitive
axes which can be achieved by controlling the ``non-sensitive''
degrees of freedom of the test masses and the position and attitude of
the spacecraft. The measurement of the test mass position is provided by a capacitive
readout system, augmented in the sensitive axes by the measurement
provided by the test mass interferometer.

The DRS can claim substantial heritage from LISA Pathfinder, as the
gravitational reference sensor will be identical and the
micro-Newton thrusters and the control law will be similar, requiring
adaptation to the larger
mass of the LISA spacecraft and the different geometry of the test
mass arrangement. Similarly, lifetime requirements for the propulsion
system are more stringent as LISA's nominal life-time is 5 years
compared to the 6 month of LISA Pathfinder.

\subsection{Environmental requirements}
\label{sec:envir-requ}

As gravitational forces cannot be shielded, the mass distribution of
the spacecraft can cause significant acceleration on the test
mass, both through direct gravitation as well as through gravity
gradients.

To be able to compensate both mass imbalance and gravity gradient, the
mass distribution on the spacecraft has to be known accurately enough
to be able to use the correct amount of compensation mass in the
correct position. On ground, a measurement of the self-gravity is not feasible
to the precision necessary for LISA, verification of the self-gravity
relies on analyses. Such analyses on self-gravity and the design of
compensation masses has been successfully performed on LISA Pathfinder
\cite{armano_gravitational_2005,brandt_lisa_2005} and the
corresponding analysis tools have been developed for LISA
\cite{merkowitz_structural_2004,merkowitz_self-gravity_2005}. 

Of similar importance is the magnetic cleanliness of the spacecraft,
as magnetic fields can cause a non-gravitational acceleration of the test
mass coupling to its non-zero magnetic susceptibility. In addition, they
create an acceleration noise as soon as the test mass carries an
electric charge. As neither the magnetic susceptibility of the
test mass nor its electric charge can be controlled to be
precisely zero, strict magnetic cleanliness has to be enacted, requiring the
use of non-magnetic materials in the vicinity of the inertial sensor.

Temperature fluctuations at the inertial sensor have the potential to
cause acceleration noise, as they will cause a time varying gas
pressure in the electrode housing leading to varying radiometer
effect. The thermal variations allowed are at the level of \SI[per=slash]{e-5}{\kelvin\per\rthz}

\subsection{Gravitational Reference Sensor}
\label{sec:grav-refer-syst}

The gravitational reference sensor (GRS) forms a crucial part of the
LISA mission and is one of the major components of the disturbance
reduction system, providing it with the data necessary to keep the
spacecraft in a (nearly) gravitational orbit. 

\begin{figure}
  \centering
  \newlength\figheight
  \newlength\figwidth
  \newbox{\tmpbox}
  \newbox{\tmpboxi}

  \savebox{\tmpbox}{\includegraphics[width=.53\linewidth]{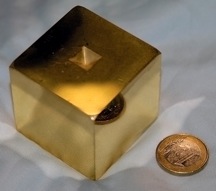}}
  \settoheight{\figheight}{\usebox{\tmpbox}}
  \settowidth{\figwidth}{\usebox{\tmpbox}} 
  \savebox{\tmpboxi}{\includegraphics[keepaspectratio=true,height=20cm,width=\figwidth]{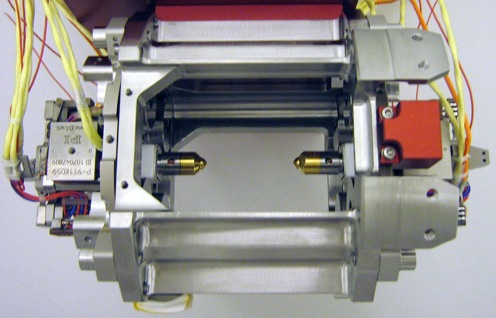}}

  \usebox{\tmpbox}
  \hfill
  \includegraphics[keepaspectratio=true,height=\figheight,width=20cm]{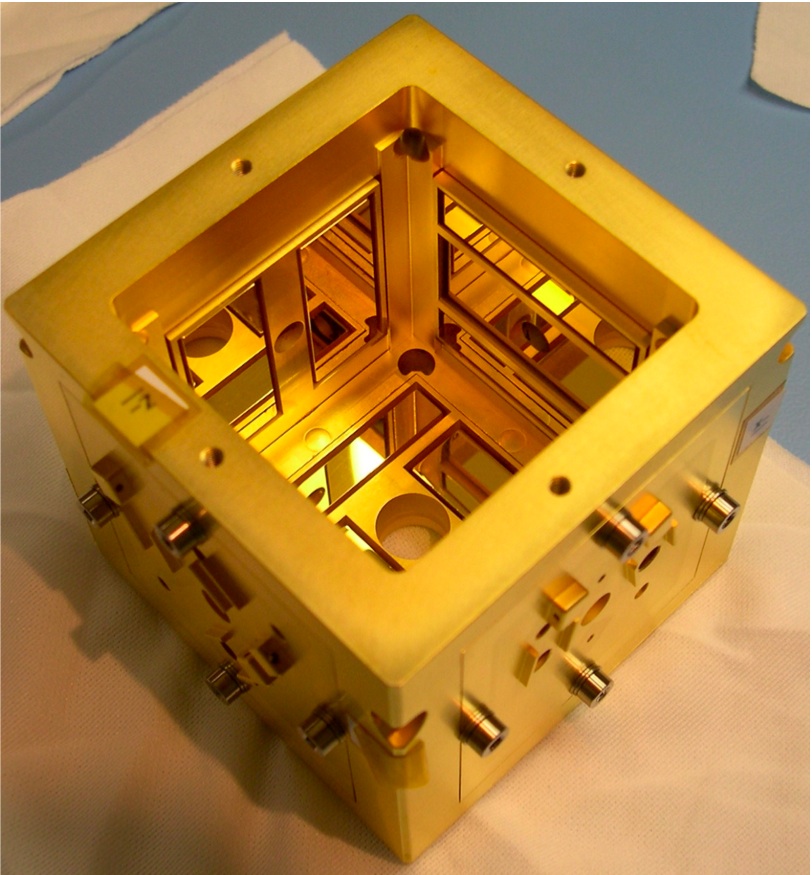}

  \vskip\baselineskip
  \settoheight{\figheight}{\usebox{\tmpboxi}}
  \includegraphics[keepaspectratio=true,height=\figheight,width=20cm]{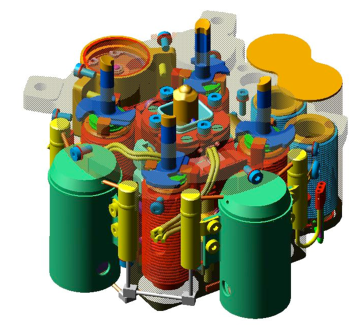}    
  \hfill
  \usebox{\tmpboxi}

  \caption{Upper left: LISA Pathfinder test mass made of a metallic
    mixture of \SI{75}{\percent} gold and \SI{25}{\percent} platinum,
    coated with gold. The prism-like impression at the centre of the
    top face takes the plunger of the caging system and allows for
    centring of the test mass, the chamfered corners accept the
    fingers during launch lock. Edges are chamfered to prevent damage
    during caging. Upper Right: Electrode Housing. Lower left:
    Schematic drawing of the bottom half of the caging mechanism, with
    the central plunger in the centre and the four hydraulically actuated
    fingers that grab onto the corners of the test mass. Lower right:
    Flight model of the caging mechanism (side view, courtesy of
    Thales Alenia Space) with both plungers visible.}
\label{fig:pm}
\end{figure}

The GRS comprises the test mass (\fref{fig:pm}, upper left), enclosed
in a housing (\fref{fig:pm}, upper right) that contains the electrodes
needed for the capacitive readout of the test mass position. The
electrodes are arranged in such a way that all relevant degrees of freedom can be
capacitively measured \cite{weber_position_2002,
  carbone_achieving_2003,stanga_ground_2006}. The GRS further contains the
launch lock mechanism (\fref{fig:pm}, lower row) and the charge
control system.

Forcing of the test mass to control its orientation and position in
the non-sensitive directions is achieved by applying AC voltages to
the electrodes; the unavoidable cross-coupling of the actuation from
non-sensitive directions into the sensitive directions has to be as small as possible (on
the order of \num{e-3}) to avoid ``leakage'' into the sensitive axis
that causes acceleration noise of the test mass. Knowledge of the correct cross-coupling
coefficients and an effective diagonalisation of the control matrix is
an important task during on-orbit commissioning of the instrument. 

Other important noise sources in the GRS to consider comprise the
electrostatic coupling between the test mass and the housing (and
ultimately the spacecraft) due to
the capacitive sensing, the \emph{stiffness} of the sensor, that
feeds the the noise of the micro-newton thrusters back into the GRS 
\cite{carbone_characterization_2005}; forces induced by thermal gradients,
such as thermal radiation pressure, or asymmetric outgassing
\cite{carbone_thermal_2007}; random charging processes
\cite{shaul_evaluation_2005,shaul_unwanted_2004}; and gas damping. 

The patch field effect, caused by spatial (and temporal) variation of
the work function, can be a major source of noise to drag-free sensors
\cite{everitt_gravity_2008}. The work function of the test mass
contributes to stray DC electrostatic fields that couple to the
time-varying charge of the surrounding electrode housing (and vice
versa), introducing both forcing and sensing noise. 

A technique to measure the stray DC field imbalances has been proposed
\cite{Weber_compensation_2007} and experimentally verified
\cite{carbone_achieving_2003} that simulates a sinusoidally varying charge on
the test mass by applying a dither voltage to selected
electrodes. Using this method, the 
average bias voltage that results from the spatial variation of the
work function can be suppressed by a factor of about 100 by applying a
DC compensation voltage, resulting in a reduction of the respective
acceleration noise to levels negligible for LISA.

The GRS is a direct heritage from the LTP experiment on LISA
Pathfinder; a detailed review on the working principle of the GRS can
be found in \cite{dolesi_gravitational_2003}. An extensive
ground-testing campaign evaluating the performance and the noise
sources on in the GRS employing a low-frequency torsion pendulum is
under way and results and more detailed descriptions of noise sources
and their effect can be found in
\cite{carbone_torsion_2006,hueller_measuringlisa_2005,carbone_characterization_2005,carbone_achieving_2003};
requirements on the sensor and the environmental conditions are
summarised in \tref{tab:is_req}.

\begin{table}
  \centering
  \caption{Summary of the environmental and performance requirements on the DRS.}
  \label{tab:is_req}
  \begin{indented}
  \item[]\begin{tabular}{@{}p{1.5in}l} \br
      Condition                              & Requirement                                               \\
      \mr 
      Acceleration                                                                                       \\
      \multicolumn{1}{r}{DC}                 & \SI{3e-9}{\meter\per\second\squared}                      \\
      \multicolumn{1}{r}{residual variation} & \SI[per=slash]{3e-15}{\meter\per\second\squared\per\rthz} \\
                                                                                                         \\
      Capacitive readout noise                                                                           \\
      \multicolumn{1}{r}{Displacement (sensitive axis)}       & \SI{1.8}{\nano\meter\per\rthz}                            \\
      \multicolumn{1}{r}{Displacement (non-sensitive axis)}       & \SI{3.0}{\nano\meter\per\rthz}                            \\
      \multicolumn{1}{r}{Rotation}           & \SI{200}{\nano\rad\per\rthz}                              \\
                                                                                                         \\
      Forcing noise                                                                                      \\
      \multicolumn{1}{r}{Sensitive axis}     & \SI{2e-15}{\meter\per\second\squared\per\rthz}            \\
      \multicolumn{1}{r}{Non-sensitive axes} & \SI{3e-14}{\meter\per\second\squared\per\rthz}            \\
      \multicolumn{1}{r}{Rotation}           & \SI{7.3e-13}{\radian\per\second\squared\per\rthz}         \\

      \\
      
      Thermal variation across the sensor               & \SI[per=slash]{e-5}{\kelvin\per\rthz}                     \\
                                                                                                                   \\
      Magnetic field                                   &                                                           \\
      \multicolumn{1}{r}{DC field}                     & \SI{4e-6}{\tesla}                                         \\
      \multicolumn{1}{r}{DC gradient}                  & \SI{e-6}{\tesla\per\meter}                                \\
      \multicolumn{1}{r}{Variation}                    & \SI{72e-9}{\tesla\per\rthz}                               \\
      \multicolumn{1}{r}{Variation of gradient}        & 
      \SI{25e-9}{\tesla\per\meter\per\rthz}                                                                        \\
                                                                                                                   \\
      Charge on test mass                                     & \num{e7} electron charge                                  \\
      Absolute position of Test mass inside electrode housing & $\SI[per=slash]{1.5e-9}{\meter\per\rthz}\times\uff[8]$    \\
      \br
    \end{tabular}
  \end{indented}  

\end{table}

\subsubsection{Test mass and housing}
\label{sec:proof-mass}

The test mass is a cube made of an alloy of about 
\SI{75}{\percent} Au  and \SI{25}{\percent} Pt with a mass of
\SI{1.96}{\kilo\gram} and dimension
$\SI{46}{mm}\times\SI{46}{mm}\times\SI{46}{mm}$. 

The mixing ratio of the two metals is chosen so that the magnetic
susceptibility $\chi$ can be made very small
\cite{budworth_thermal_1960,silvestri_volume_2003}. As the
susceptibility depends on the mixing ratio and the manufacturing
process, a small residual magnetic susceptibility of
$\chi\approx\SI{-2e-5}{}$ remains in the test mass, requiring a
certain amount of magnetic cleanliness of the whole spacecraft that
prohibits the use of ferro-magnetic materials in the vicinity of the
GRS.

The surface of the test mass is coated with a thin layer of gold that
provides reflectivity for the laser light of the local
interferometer. In addition gold proves to be the material of choice
to minimise the patch field effect.

The test mass is surrounded by a housing that contains the electrodes
for the capacitive sensing and actuating. The housing is slightly
larger that the test mass, with the gap between the
test mass and the electrodes measures between \SI{3}{mm} and
\SI{4}{mm} provides a further reduction of the patch field effect, as
the noise forces decrease with  the distance between test mass and
electrodes. An additional benefit of the large gaps is a reduction of
the dissipation due to gas flow around the test mass. 

The electrode housing admits the the fingers and the plunger of the
launch lock and repositioning mechanism (see
\sref{sec:launch-lock-repos}) in the $Z$ surfaces and the laser
of the test mass interferometer through a hole in the $X$ surface.

The electrodes are made from a gold-coated sapphire substrate,
surrounded by a molybdenum guard ring; the electrode housing structure
is made from molybdenum as well.

The physical properties of test mass and housing are summarised in \tref{tab:tmhousing_req}.
 
\begin{figure}
  \centering
  \includegraphics[height=.3\textheight, keepaspectratio]{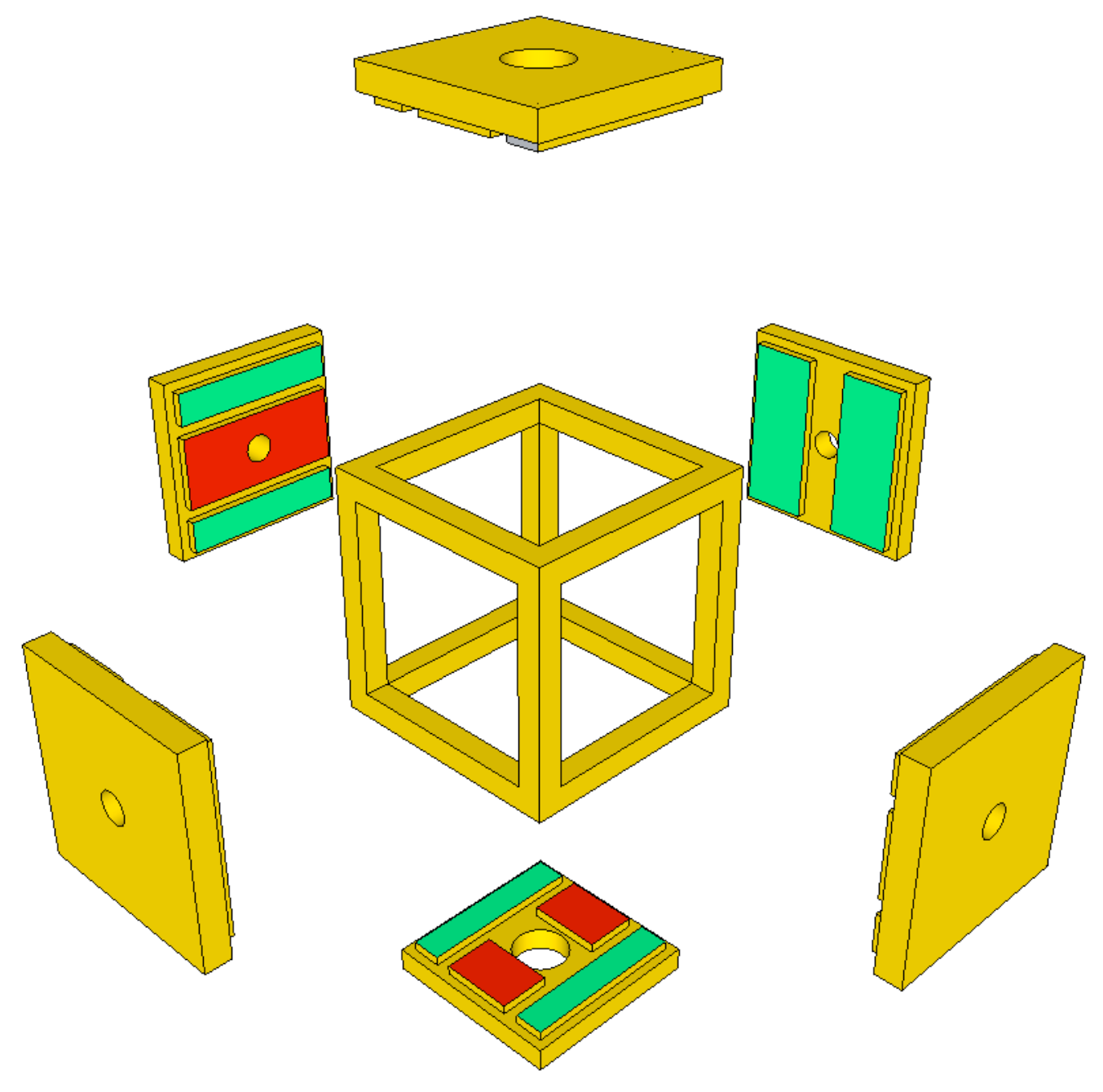}
  \caption{Exploded schematic view of the electrode housing, showing the
    structure of the electrode housing and the electrodes mounted on
    the inner faces.}
  \label{fig:EH_exploded_view}
\end{figure}

\begin{figure}
  \centering

  \svnidlong 
{$HeadURL: file:///Users/ojennric/SVN/Paper/LISA_Technology/electrode_configuration.tex $} 
{$LastChangedDate: 2009-03-09 17:29:26 +0100 (Mon, 09 Mar 2009) $} 
{$LastChangedRevision: 33 $} 
{$LastChangedBy: ojennric $} 
\svnid{$Id: electrode_configuration.tex 33 2009-03-09 16:29:26Z ojennric $} 

\begin{tikzpicture}[%
,key/.style={font=\footnotesize}%
,electrode/.style={green, border green}%
] 


\begin{scope}[scale=0.8]

  \draw (20.2mm, 48mm) node [key] {$X$ face};

  \draw [->,thick] (-3mm,15.2mm) -- node [key,left] {$z$} +(0,10mm) ;
  \draw [->,thick] (15.2mm, -3mm) -- node [key,below] {$y$} +(10mm,0 ) ;  

  \draw (0,0) rectangle (40.4mm, 40.4mm);

  \draw [shift={(1.2mm,1.2mm)}] (0,0) rectangle (15.7mm, 38.0mm);
  \fill [shift={(2.2mm,2.2mm)},green] (0,0) rectangle (13.7mm, 36.0mm);

  \draw [shift={(39.2mm,1.2mm)}] (0,0) rectangle (-15.7mm, 38.0mm);
  \fill [shift={(38.2mm,2.2mm)},green] (0,0) rectangle (-13.7mm, 36.0mm);

  \draw (20.2mm,20.2mm) circle (3.0mm);

\end{scope}

\begin{scope}[xshift= 41mm, scale=0.8]

  \draw (21.3mm, 48mm) node [key] {$Y$ face};

  \draw [->,thick] (-3mm,16.3mm) -- node [key, left] {$x$} +(0,10mm) ;
  \draw [->,thick] (16.3mm, -3mm) -- node [key, below] {$z$} +(10mm,0 ) ;  

  \draw (0,0) rectangle (42.6mm, 42.6mm);

  \draw [shift={(1.2mm,1.2mm)}] (0,0) rectangle (9.1mm, 40.2mm);
  \fill [shift={(2.2mm,2.2mm)},green] (0,0) rectangle (7.1mm, 38.2mm);

  \draw [shift={(41.4mm,1.2mm)}] (0,0) rectangle (-9.1mm, 40.2mm);
  \fill [shift={(40.4mm,2.2mm)},green] (0,0) rectangle (-7.1mm, 38.2mm);

  \draw [shift={(12.3mm,1.2mm)}] (0,0) rectangle (18mm, 40.2mm);
  \fill [shift={(13.3mm,2.2mm)},red] (0,0) rectangle (16mm, 38.2mm);

  \draw [fill=white] (21.3mm,21.3mm) circle (3.0mm);

\end{scope}

\begin{scope}[xshift=83mm, scale=0.8]

  \draw (20.7mm, 48mm) node [key] {$Z$ face};

  \draw [->,thick] (-3mm,15.2mm) -- node [key, left] {$x$} +(0,10mm) ;
  \draw [->,thick] (15.2mm, -3mm) -- node [key, below] {$y$} +(10mm,0 ) ;  

  \begin{scope}[rotate around={90:(20.7mm,20.7mm)}]

    \draw (0,0) rectangle (41.4mm, 41.4mm);

    \draw [shift={(1.2mm,1.2mm)}] (0,0) rectangle (39mm, 8.5mm);
    \fill [shift={(2.2mm,2.2mm)},green] (0,0) rectangle (37mm, 6.5mm);

    \draw [shift={(1.2mm,40.2mm)}] (0,0) rectangle (39mm, -8.5mm);
    \fill [shift={(2.2mm,39.2mm)},green] (0,0) rectangle (37mm, -6.5mm);

    \draw [shift={(1.2mm,11.7mm)}] (0,0) rectangle (13.2mm, 18mm);
    \fill [shift={(2.2mm,12.7mm)},red] (0,0) rectangle (11.2mm, 16mm);

    \draw [shift={(40.2mm,11.7mm)}] (0,0) rectangle (-13.2mm, 18mm);
    \fill [shift={(39.2mm,12.7mm)},red] (0,0) rectangle (-11.2mm, 16mm);

    \draw [fill=white] (20.7mm,20.7mm) circle (6.0mm);
  \end{scope}
\end{scope}

\begin{pgfonlayer}{background} 

\end{pgfonlayer}
\end{tikzpicture}
  \caption{Placement of the electrodes onto the inner surfaces of the
    electrode housing. Control electrodes are given in green (light grey),
    injection electrodes in red (dark grey). The central holes in the $X$- and
    $Y$-faces admit the laser, the central hole in the $Z$ face admits
  the plunger. The electrodes differ slightly in overall size on the
  different faces.}
  \label{fig:electrode_conf}
\end{figure}
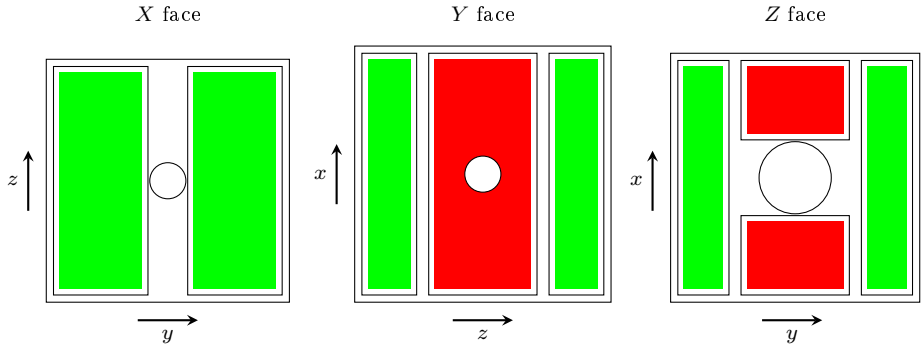

\begin{table}
  \centering
  \caption{Summary of the physical parameters of the test mass and the
    housing.}
  \label{tab:tmhousing_req}
  \begin{indented}
  \item[]\begin{tabular}{@{}p{1.5in}l} \br
          Element                                 & Property                                        \\
  \mr
      Test mass                                                                                     \\
      \multicolumn{1}{r}{Size}                    & $\SI{46}{mm}\times\SI{46}{mm}\times\SI{46}{mm}$ \\
      \multicolumn{1}{r}{Material}                & gold-coated AuPt (\SI{75}{\percent} Au,
      \SI{25}{\percent} Pt )                                                                        \\
      \multicolumn{1}{r}{Mass}                    & \SI{1.96}{kg}
      \\
       \multicolumn{1}{r}{Magnetic susceptibility} & $|\chi|\leq \num{2e-5}$ \\
                                                                                                    \\
      Housing                                                                                       \\
      \multicolumn{1}{r}{Material}                & gold-coated molybdenum                          \\
    \multicolumn{1}{r}{Gaps Electrodes/Test mass} & \SI{4}{mm} (x), \SI{2.9}{mm} (y), \SI{3.5}{mm} (z)                                                         \\
                                                                                                    \\
      Electrodes                                                                                    \\
      \multicolumn{1}{r}{Material}                & 
      gold-coated sapphire                                                                          \\
      \multicolumn{1}{r}{Size and Arrangement}    & see \fref{fig:electrode_conf}                   \\
      \br
  \end{tabular}
\end{indented}
\end{table}
\subsubsection{Capacitive sensing}
\label{sec:capacitive-sensing}

The capacitive sensing of the test mass position is designed to
minimise the disturbance of the test mass while at the same time
yielding a measurement with noise levels of
\SI[per=slash]{1.8}{\nano\meter\per\rthz}  for the sensitive axis. Six opposing
pairs of electrodes form a differential capacitive-inductive bridge
with a resonance frequency of about $\omega_0 =
2\pi\times\SI{100}{\kilo\hertz}$. Combinations of the obtained signals
yield all displacements and rotations. In order to apply the AC bias
to the test mass, injection electrodes are placed on the $+Z$ and $-Z$
as well as on the $+Y$ and $-Y$ surfaces of the electrode housing (\fref{fig:electrode_conf}). 
The capacitive sensing achieves a
sensitivity of $\SI[per=slash]{2}{\nano\meter\per\rthz}$ in displacement and
$\SI[per=slash]{200}{\nano\radian\per\rthz}$ in rotation
\cite{carbone_upper_2007} in ground tests, matching or exceeding the
requirements for LISA.

The details of the sensor design and a detailed discussion of the
noise can be found in \cite{Cavalleri:2001:pdp}.

\subsubsection{Launch lock and repositioning}
\label{sec:launch-lock-repos}

The relatively large gaps make it necessary that the test mass is
held fixed during launch by the \emph{caging mechanism} to avoid damage to the
test mass or the electrode housing due to the vibrations present
during launch. During launch lock, eight
hydraulically actuated fingers connect to the eight corners of the
cubical test mass, each pushing with a force of \SI{1200}{N} to keep
the test mass securely in place (see \fref{fig:pm}, lower left
panel for a drawing of the caging mechanism). 

Releasing the test mass from the launch lock requires to break the
adhesion present between the fingers and the surface of the test
mass. The necessary force to break the adhesion can be up to
\SI{10}{N} per finger (on the order of \SI{1}{\percent} of the load),
so that without a way to push the test mass off the fingers, it would
remain stuck to the launch lock. In addition, the residual momentum
of the test mass after release needs to be smaller than
\SI{e-5}{\newton\second} for the electro-static actuator to be able to
slow down and centre the test mass in the electrode housing. 

A multi-stage approach for the release of the test mass has been taken:
To overcome the adhesion between the fingers and the test mass, two
piezo-driven plungers, acting centrally on the $+Z$ and $-Z$ surface
of the test mass, respectively, are used to push the test mass off the
fingers. The $Z$ surfaces of the test mass have pyramidal/conical
indentations to allow for an auto-centring and auto-aligning of the
test mass during engagement of the plungers (see \fref{fig:pm},
upper left panel).

As the plungers push with up to \SI{40}{N} into the indentations, an
adhesion force of about \SI{0.5}{N} will have to be overcome when
attempting to retract the plungers. For that purpose, the plungers
accommodate a release tip at their end (much like a retractable
ball-point pen) that can be pushed out by a piezo-electric element to
deliver the necessary force. The remaining adhesion, still too large
to be overcome by the electro-static actuator \cite{benedetti:828}, is
then broken using the inertia of the test mass by quickly retracting
the plunger, leaving the test mass with residual momentum below the
specified \SI{e-5}{\newton\second}.

After launch, only the plungers are employed to grab and position the
test mass during spacecraft safe mode or any other circumstance that
makes it necessary to re-position the test mass.

The breaking of the adhesion between plungers and test mass has been
the topic of intense ground-based testing, showing the feasibility of
a test mass release within the required limits of the transferred
momentum \cite{Bortoluzzi:2009:tmi}.

\begin{figure}
  \centering
  \input{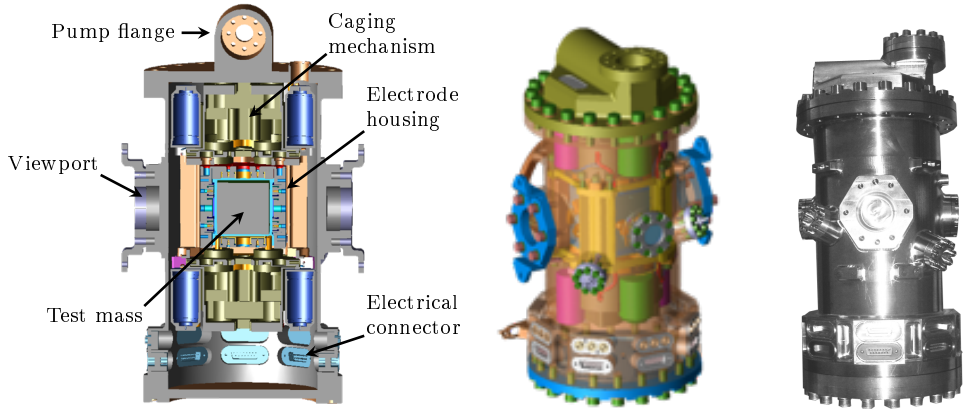}
  \caption{Left: Schematic drawing of the vacuum housing for the
    inertial sensor of LPF. Centre: Schematic drawing, external
    view. Right: Flight model of the vacuum housing for the
    inertial sensor of LPF. The inertial sensor for LISA is foreseen
    to be identical to the sensor used in LISA Pathfinder.}
\label{fig:is_housing}
\end{figure}

\subsubsection{Vacuum system}
\label{sec:vacuum-system}

It is a peculiarity of LISA (and LISA Pathfinder) that despite the
fact that the mission will operate in interplanetary space, it needs
to carry a vacuum system. The residual gas pressure due to outgassing in the inertial
sensor is
too high for the gravitational reference sensor to tolerate, as it
creates spurious noise due to the radiometric effect
\cite{carbone_thermal_2007} and through gas damping. Therefore, a vacuum system
(\fref{fig:is_housing}), pumped by a getter pump and encapsulating the
test mass, the electrode housing and ancillary structures is foreseen,
maintaining a pressure of \SI{e-8}{mbar}. Due to the critically
of ferro-magnetic materials that close to the test masses, this vacuum
system has been made entirely from titanium.

\subsubsection{Charge control system}
\label{sec:charge-contr-syst}

As forcing of the test mass depend to a large degree on
electrostatic forces, the electric charge of the test mass has to be
controlled as any fluctuation in the charge of the test mass will
give rise to a fluctuation in force, hence acceleration noise. 

Charging of the test mass mainly occurs when secondary particles
created by interaction of either protons or $\alpha$-particles from
cosmic radiation with the spacecraft materials hit the test mass
\cite{jafry_electrostatic_1997,sumner_description_2004}. The charging
rates incurred are on the order of \SI{50}{e/s}.

A standard way to discharge test masses in similar setups is to
connect a thin wire of conductive material to the test mass
\cite{touboul_electrostatic_1996}, however, such a mechanical
connection introduces spurious accelerations and proves to be too
noisy for the LISA requirements. Another well proven way to remove
surface charge is through the photo-electric effect. In the case of
LISA, UV light will be used to irradiate test mass and electrode
housing, removing surface charges from electrodes and test mass.

The charge control system for LISA is based on heritage from the LISA
Pathfinder mission \cite{schulte_charge-management_2006}, which itself
is based on the charge control system flown on the GP-B mission, and
whose functionality and performance of the discharging system has been
demonstrated \cite{wass_testing_2006}. The LPF charge control system
consists of six mercury discharge lamps, producing UV light at
\SI{254}{nm} coupled into optical fibres and brought to the test mass
(2 lamps) and electrode housing (1 lamp). An identical setup controls the
charge of the second test mass, bringing the number of lamps up to the
total of six. 

Due to the reflectivity of both the electrodes and the test mass,
light shone on any surface will eventually reach most of the other
surfaces as well and release electrons, so that the discharge rate is
determined by the net current between electrodes and test mass.
The polarity of the discharge is controlled by the digitally
controlled output power of the UV
lamps received by the test mass and the electrode housing,
respectively, and can be further controlled by applying bias voltages
to the electrodes.

Operationally, the discharging can occur episodic or continuously,
depending on the observed charge rate. The charge itself is measured
by applying an AC voltage to the electrodes and measure the ensuing
displacement of the test mass \cite{sumner_description_2004, schulte_charge-management_2006}.

For LISA the development of UV LED \cite{sun_led_2006} opens the
possibility to replace the mercury discharge lamps with LED requiring
less power and having less mass.

\subsection{Micro-newton thrusters}
\label{sec:micr-thrust}

\begin{table}
\caption{Summary of the micro-Newton thruster requirements.}
  \label{tab:mnt_req}
  \begin{indented}
    \item[]
      \begin{tabularx}{\linewidth}{@{}>{\setlength\hsize{.5\hsize}}X>{\setlength\hsize{1.5\hsize}}X}
        \br
        &Requirement\\
        \mr
        Minimum thrust & $\SI{0.3}{\micro\newton}$\\
        Maximum thrust & $\SI{100}{\micro\newton}$\\[1.5ex]
        Thrust resolution & \SI{0.3}{\micro\newton} \\
        Thrust noise& $\SI{0.1}{\micro\newton\per\rthz}\times\uff[10]$ \\
        Lifetime&\SI{55000}{hours}\\
        Specific impulse&\SI{4000}{s}\\
        Total impulse&\SI{8300}{\newton\second} per thruster\\
        \br
    \end{tabularx}
  \end{indented}  
\end{table}

\begin{figure}
  \centering
  \includegraphics[width=0.4\linewidth]{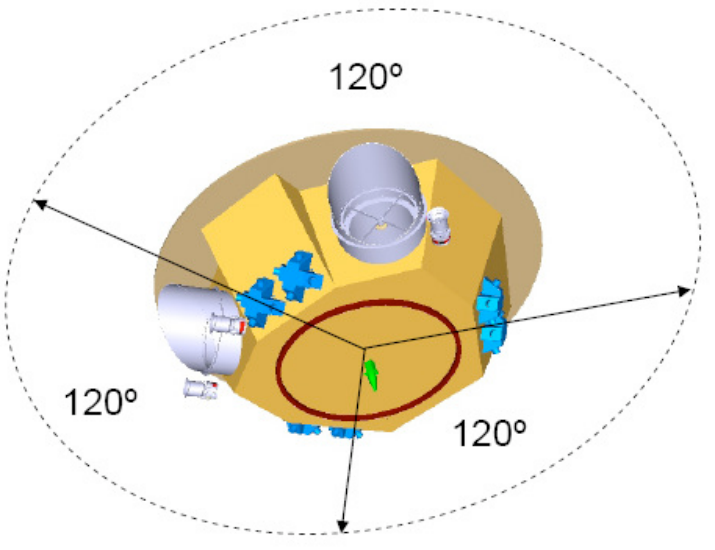}\hfill
  \includegraphics[width=0.4\linewidth]{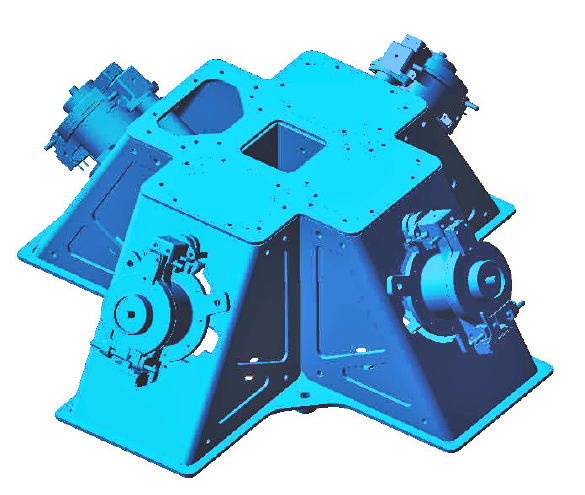}
  \caption{Geometrical layout of the micro-newton thrusters on the LISA spacecraft (left) and FEEP cluster module 
    geometry (right). Drawing courtesy of EADS Astrium}
\label{fig:feep_arrangement}
\end{figure}

The micro-newton thrusters are a key technology for LISA, as they are
providing the fine attitude and position control for the drag free
flight. LISA will employ three clusters of four thrusters each,
situated on the outside of the spacecraft, separated by
\SI{120}{\degree} (see~\fref{fig:feep_arrangement}, allowing to control all
degrees of freedom of the spacecraft.

The thrusters are controlled by the DRS and operate continuously during
science operation. The main thrust is used to counteract the solar
radiation pressure that amounts to about \SI{10}{\micro\newton} per
relevant thruster, the largest external force on the spacecraft. The
thrust noise is required to be smaller than \SI{0.1}{\micro\newton} at
high frequencies (relaxing to lower frequencies, see
\tref{tab:mnt_req}) in order to keep the motion of the spacecraft with
respect to the test masses as small as possible. 

Two different propulsion systems currently meet the LISA requirements
on thrust and thrust noise, both based on field emission ionisation of
the propellant. A colloid micro-newton thruster (CMNT) developed in the US
and the European field emission electric propulsion system
(FEEP). Both thruster systems will be flown on LISA Pathfinder
\cite{mcnamara_lisa_2008} to
demonstrate the technology and to assess the on-orbit performance.

The US CMNT uses a colloidal liquid as a propellant. Small droplets of
the colloid are ionised through field emission, accelerated in an
electrical field and ejected from the thruster. The thrust is over a
wide range proportional to the acceleration voltage and can be
controlled with the required precision. The CMNT has shown a
capability of \SI{15}{\micro\newton} thrust with a noise well below the
requirement and has successfully passed a \SI{3400}{hour} life time
test.

The ESA FEEP development programme advanced two different technologies,
based on indium \cite{genovese_new_2004,tajmar_indium_2004,steiger_micronewton_2000,fehringer_indium_1998,rdenauer_indium_1997}
and caesium  \cite{marcuccio_experimental_1998} with different
geometrical setups for the field emission. The advantage over
the CMNT lies in the much higher specific impulse of the FEEP, as single
ions instead of charged droplets are accelerated and ejected, yielding
a better charge to mass ratio.

In the In-FEEP capillary forces push the indium to the tip of a needle
needle where it is ionised, whereas the Cs-FEEP employ a narrow slit
in which the caesium then forms narrow cones due to the electric
field, emitting ions from the tip of the cones, producing higher
thrust than the In-FEEP due to the multiple emitters.

The Cs-FEEP have been chosen as the baseline technology to fly on
LISA~PF and have demonstrated the required thrust, noise and
resolution and have performed endurance testing with a total impulse
in excess of \SI{800}{\newton\second} in over \num{3000} hours of
operation; a campaign to measure the thrust directly on a nano-balance
is ongoing, aimed at verifying the thrust measurements obtained from
monitoring the electrical current produced by the thrusters.

While the currently demonstrated lifetime and total impulse of both
micro-newton thruster systems is sufficient for the LISA Pathfinder
mission, a higher lifetime and total impulse has to be demonstrated for
LISA. The additional life-time testing is part of ESA's technology
development programme for LISA.


\svnidlong 
{$HeadURL: file:///Users/ojennric/SVN/Paper/LISA_Technology/Conclusion.tex $} 
{$LastChangedDate: 2009-06-15 11:26:41 +0200 (Mon, 15 Jun 2009) $} 
{$LastChangedRevision: 50 $} 
{$LastChangedBy: ojennric $} 
\svnid{$Id: Conclusion.tex 50 2009-06-15 09:26:41Z ojennric $} 

\section{Conclusions}
\label{sec:conclusions}

LISA is a the first space mission dedicated to gravitational waves and
will deliver a rich scientific payoff. The technology that is employed
by LISA has seen significant advances in the last couple of years, not
least to the technology development programme for the technology
precursor mission LISA Pathfinder. 

The main technology areas of LISA are the interferometric measurement
system, comprising the laser, the telescope, the optical bench and the
phase measurement; and the disturbance reduction system, consisting of
the gravitational reference sensor with test mass, housing and caging
mechanism, the control law, and the micro-Newton thrusters.

LISA can claim significant heritage
form LISA Pathfinder in the DRS, where almost all of the technologies
will be flight-tested during the LPF mission: the gravitational
reference sensor, the micro-newton thrusters, and the charge-control
system. While the control law requires adaption due to the different
mass and moments of inertia of LISA and due to the fact the LISA has
to keep two test masses in free fall condition, the principle of a
drag-free control system able to operate at the required levels of
noise will be demonstrated.

The technology needed for the interferometric measurements benefits
less from the LPF mission, however, important issues such as the
construction of an ultra-stable optical bench and the design principles
of the interferometer are part of the technology that LPF will
demonstrate.

The remaining technology items, such as the phase measurement for
LISA, the telescope, and the high power laser system are subject to an
ongoing technology programme conducted by ESA and NASA as well as the
participating national space agencies. 


\svnidlong 
{$HeadURL: file:///Users/ojennric/SVN/Paper/LISA_Technology/Acknowledgement.tex $} 
{$LastChangedDate: 2009-03-25 10:42:53 +0100 (Wed, 25 Mar 2009) $} 
{$LastChangedRevision: 34 $} 
{$LastChangedBy: ojennric $} 
\svnid{$Id: Acknowledgement.tex 34 2009-03-25 09:42:53Z ojennric $} 

\section{Acknowledgement}
\label{sec:acknowledgement}

The author would like to thank P~McNamara, G~Heinzel, W~Weber and D
Nicolini for providing many details on the technology and useful
feedback, K~Danzmann and S~Vitale for helpful comments, and EADS
Astrium GmbH, in particular D~Weise and P~Gath, for providing
high-quality graphics for the optical bench, the optical assembly and
the micro-propulsion system.



\printbibliography

\end{document}